\documentclass[acmsmall]{acmart}

\setcopyright{cc}
\setcctype{by}
\acmDOI{10.1145/3832219}
\acmYear{2026}
\acmJournal{PACMSE}
\acmVolume{3}
\acmNumber{ISSTA}
\acmArticle{ISSTA128}
\acmMonth{10}
\acmSubmissionID{issta26main-p1189-p}
\received{2026-01-30}
\received[accepted]{2026-06-25}

\usepackage{algorithm}
\usepackage{algpseudocode}

\usepackage{tabularray}
\usepackage{booktabs}
\usepackage{multirow}
\usepackage{adjustbox}
\usepackage{siunitx}
\usepackage{array}  
\usepackage{tabularx}
\usepackage{caption}
\usepackage{colortbl}
\usepackage{xcolor}
\usepackage{makecell}
\usepackage{libertine}
\usepackage{xspace}
\usepackage{arydshln}
\usepackage{enumerate}
\usepackage{enumitem}
\usepackage{pifont}
\usepackage{amsmath}
\usepackage{tikz}
\usepackage[normalem]{ulem}
\usepackage{wrapfig}

\newcommand{\blackcircled}[1]{%
  \tikz[baseline=(char.base)]{%
    \node[shape=circle, fill=black, inner sep=1pt, text=white] (char) {#1};%
  }%
}

\newcommand{\MCRbench}{\textsc{MCR-Bench}\xspace}
\usepackage[most]{tcolorbox}

\newcommand{\boxmargin}{2mm}
\tcbuselibrary{most, skins, breakable, theorems}
\newtcolorbox{myboxa}[2][]{
    colback=gray!10!white,
    colframe=black, enhanced,
    attach boxed title to top left={yshift=-2mm,xshift=5mm},
    title=#2,#1
}
\newtcolorbox{myboxb}[2][]{
    boxsep=1pt,
    left = \boxmargin, right = \boxmargin, top = \boxmargin, bottom = \boxmargin,
    title={#2},#1
}

\newtcolorbox{myboxc}{
    colback=yellow!10!white,
    colframe=gray!50,
    arc = 0pt, outer arc = 0pt,
    boxsep=0pt, left = 3pt, right = 0pt, top = 0pt, bottom = 0pt, 
    leftrule=3pt,
    bottomrule=0pt, toprule=0pt, rightrule=0pt,
    left = \boxmargin, right = \boxmargin, top = \boxmargin, bottom = \boxmargin,
    before skip=5pt,
    after skip=5pt
}

\begin{document}

\title{From Static to Dynamic: Benchmarking Real-World Code Review with \MCRbench}

\author[D. Zheng]{Dewu Zheng}
\affiliation{%
  \institution{Sun Yat-sen University}
  \department{Zhuhai Key Laboratory of Trusted Large Language Models}
  \city{Zhuhai}
  \country{China}
}
\email{zhengdw5@mail2.sysu.edu.cn}

\author[Y. Wang]{Yanlin Wang}
\correspondingauthor
\authornote{Yanlin Wang is the corresponding author.}
\affiliation{%
  \institution{Sun Yat-sen University}
  \department{Zhuhai Key Laboratory of Trusted Large Language Models}
  \city{Zhuhai}
  \country{China}
}
\email{wangylin36@mail.sysu.edu.cn}

\author[X. Wang]{Xiwen Wang}
\affiliation{%
  \institution{Sun Yat-sen University}
  \department{Zhuhai Key Laboratory of Trusted Large Language Models}
  \city{Zhuhai}
  \country{China}
}
\email{wangxw86@mail2.sysu.edu.cn}

\author[K. Duan]{Kefeng Duan}
\affiliation{%
  \institution{Sun Yat-sen University}
  \department{Zhuhai Key Laboratory of Trusted Large Language Models}
  \city{Zhuhai}
  \country{China}
}
\email{duankf@mail2.sysu.edu.cn}

\author[H. Zhang]{Hongyu Zhang}
\affiliation{%
  \institution{Chongqing University}
  \city{Chongqing}
  \country{China}
}
\email{hyzhang@cqu.edu.cn}

\author[X. Liu]{Xilin Liu}
\affiliation{%
  \institution{Huawei Cloud Computing Technologies Co., Ltd.}
  \city{Dongguan}
  \country{China}
}
\email{liuxilin3@huawei.com}

\author[Y. Ma]{Yuchi Ma}
\affiliation{%
  \institution{Huawei Cloud Computing Technologies Co., Ltd.}
  \city{Dongguan}
  \country{China}
}
\email{mayuchi1@huawei.com}

\author[Z. Zheng]{Zibin Zheng}
\affiliation{%
  \institution{Sun Yat-sen University}
  \department{Zhuhai Key Laboratory of Trusted Large Language Models}
  \city{Zhuhai}
  \country{China}
}
\email{zhzibin@mail.sysu.edu.cn}



\begin{abstract}

In real-world software development, code review typically involves iterative interactions between developers and reviewers to improve software quality, making the process costly and time-consuming. Although recent work explores large language models (LLMs) for automated code review, most approaches oversimplify code review into a single-round, static decision task, which fails to capture the multi-round interactive nature and the complex problem-solving processes inherent in realistic review scenarios. 
To bridge this gap, we introduce \MCRbench, the first defect state-aware benchmark designed for realistic multi-round code review. \MCRbench covers five commonly-used programming languages and consists of 2,269 real-world multi-round code review tasks, each of which is annotated with fine-grained defect information and cross-round state labels. 
Each task in \MCRbench is equipped with fine-grained defect metadata (e.g., description, type, severity) alongside dynamic state annotations, capturing the complete evolutionary trajectory of a defect throughout the multi-round process.
We obtain several findings through extensive experiments on \MCRbench with mainstream LLMs.
\textbf{(1) Limited overall capability:}
experiments reveal that mainstream LLMs exhibit limited overall performance in defect detection and defect lifecycle state tracking, with performance degrading significantly as the number of interaction rounds increases;
\textbf{(2) Defect-sensitive performance:} LLMs' performance varies substantially across different defect types and severity levels, with semantically complex or low-salience defects being significantly more likely to be missed;
\textbf{(3) Underlying Failure Mechanisms:} our in-depth error analysis dissects the distinct drivers of false positives and false negatives, revealing critical weaknesses such as cross-round temporal misalignment and inadequate long-range memory.

\end{abstract}
\maketitle

\section{Introduction}

Code review stands as a cornerstone of software quality assurance~\cite{p6,p7,p8}, aiming to enhance functional correctness and long-term maintainability before code integration. While pivotal, this process is historically labor-intensive and difficult to scale~\cite{p9,p10}, which has motivated recent efforts to leverage LLMs for automated code review~\cite{ren2025hydra,li2022automating,jiang2025deep,lin2025codereviewqa,codefuse,swrbench,sun2025bitsai,chen2025understanding}.

Although existing studies have extended their code review granularity from traditional diff hunks~\cite{peng2025icodereviewer,jaoua2025combining,lu2023llama,liu2025securereviewer,hong2025retrieval} to Pull Request (PR) level~\cite{swrbench,codefuse, zhang2026sphinx,goccmen2025enhanced,cihan2025automated} to better approximate real-world development scenarios, they predominantly reduce the complex review process to a single-round, static decision task. Such simplification neglects the fact that real-world code review typically involves iterative interactions between developers and reviewers, and therefore fails to capture two critical characteristics of real-world code review: \textbf{multi-round review interactivity} and \textbf{dynamic defect evolution}~\cite{10.1007/s10664-023-10411-x,10.1007/s10664-022-10205-7,zheng2024humanevo}.

\begin{figure}[t]
    \centering
    \includegraphics[width=0.95\textwidth]{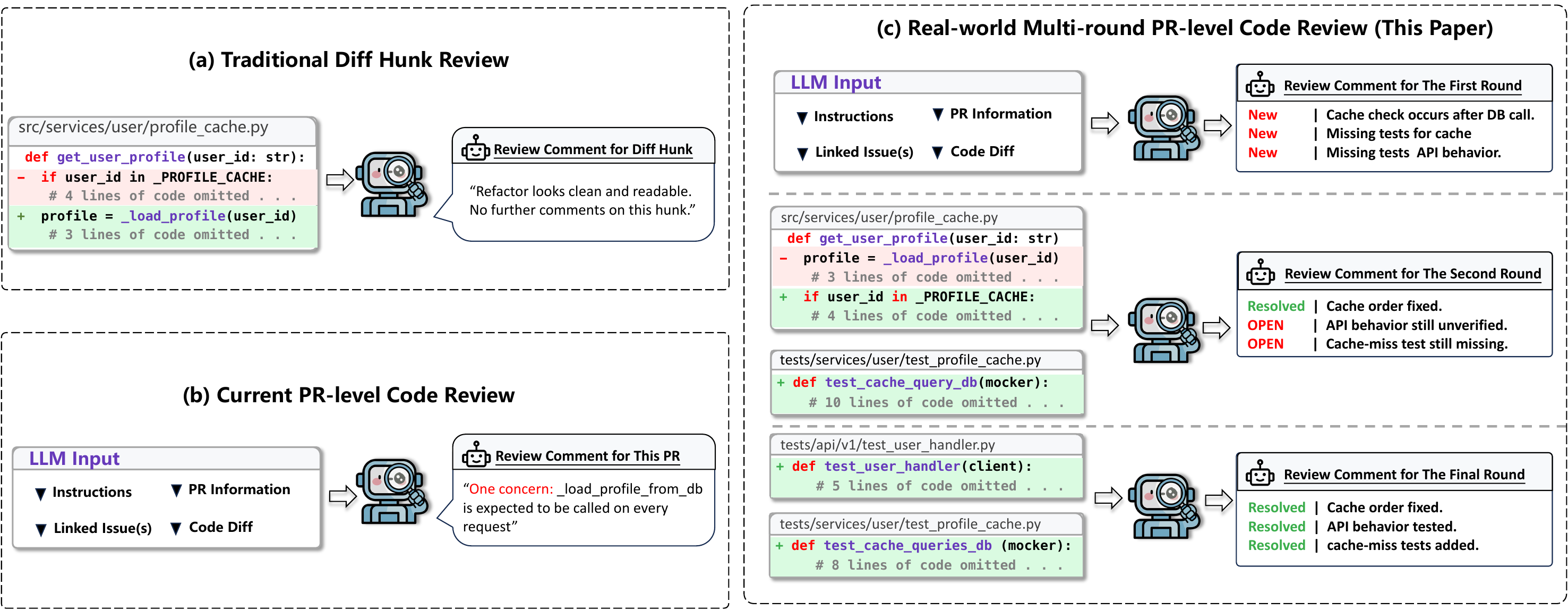}
    \caption{Comparison of traditional hunk-level review, current PR-level code review, and real-world multi-round PR-level code review (this paper).}
    \label{fig:comparison_examples}
\end{figure}

As depicted in Figure~\ref{fig:comparison_examples}, traditional review approaches (Figure~\ref{fig:comparison_examples}a) are limited to operating at the level of isolated diff hunks.
More recent PR-level benchmarks (Figure~\ref{fig:comparison_examples}b)~\cite{swrbench,codefuse,zhang2026sphinx} elevate the evaluation granularity to the PR-level to better reflect realistic review practices; however, they remain constrained to generating single-round feedback based on the initial PR state and fail to support the continued tracking and validation of the identified defects.
In practice, code reviews (Figure~\ref{fig:comparison_examples}c) typically proceed through multiple review rounds with successive commits~\cite{10.1007/s10664-022-10205-7}. Defects raised in early rounds evolve as the code changes—being resolved or remaining open for further discussion. In such dynamic settings, defect semantics depend critically on historical states and prior feedback. Without explicitly modeling multi-round interactions and state transitions, LLMs cannot reliably assess whether current changes address past review comments, limiting their effectiveness in real-world applications.

Notably, empirical evidence from real-world Gerrit projects~\cite{10.1007/s10664-021-10035-z} shows that nearly half of code changes involve multiple review rounds, with review time increasing from 0.33 days (single-round) to 5.3 days (2–6 rounds) and 31.3 days (>$6$ rounds), highlighting the critical importance of modeling multi-round interactions and defect evolution.

\textbf{\MCRbench}. To bridge the gap and advance code review automation toward more realistic interactive scenarios, we introduce \MCRbench, the first benchmark dedicated to multi-round code review, comprising 2,269 real review tasks across the five popular languages~\cite{Octoverse}: Python, Java, JavaScript, TypeScript, and C\#.
Unlike existing benchmarks that focus primarily on static comment generation, \MCRbench is designed to evaluate LLMs' ability to track defect states and maintain contextual consistency under dynamic, multi-round interactions. To this end, we annotate each review task with fine-grained ground-truth defect cards, which capture detailed defect descriptions, locations, categories, and severity levels. More importantly, we explicitly record the lifecycle state transitions of each defect across multiple review rounds (e.g., New $\rightarrow$ Open $\rightarrow$ Resolved).

\textbf{Construction Process.} 
To help ensure the quality of \MCRbench, we design a rigorous multi-stage data construction pipeline. Specifically, after performing high-quality repository selection and PR data collection, we employ an LLM-based state-aware defect annotation pipeline to address the challenge of defect tracking in multi-round code review. The pipeline follows a ``local detection first, global tracking later'' strategy: it first identifies candidate defects within individual review rounds and then performs cross-round merging and lifecycle tracking by leveraging the complete code changes and review discussions to track defect state evolution throughout the PR lifecycle. Finally, manual cross-validation is conducted to enhance the accuracy and consistency of the annotations.

\textbf{Experiments \& Findings}
We conduct an empirical evaluation of mainstream LLMs~\cite{anthropic,gpt,deepseek,gemini,glm,kimi,qwen} on \MCRbench to assess their capabilities in multi-round code review. Our evaluation focuses on defect identification and lifecycle state tracking, along with an analysis of common error patterns, yielding the following key findings.
\blackcircled{1} Limited overall capability. Current LLMs show modest performance on \MCRbench, particularly in defect identification, and even when defects are correctly identified, lifecycle state prediction remains a significant challenge. In addition, LLMs' performance degrades as the number of review rounds increases.
\blackcircled{2} Defect-sensitive performance. LLMs exhibit substantial variation in performance across defect types and severity levels, with semantically complex or low-salience defects being significantly more prone to omission.
\blackcircled{3} Underlying Failure Mechanisms: In-depth error analysis attributes the high rates of false positives and false negatives to specific cognitive deficiencies in multi-round contexts. Models suffer from inadequate cross-round memory, leading to defect forgetting and temporal misalignment, where models fail to correctly distinguish between historical, resolved, and newly introduced issues.

\noindent\textbf{Our contributions are summarized as follows:}
\begin{itemize}[leftmargin=10pt]
    \item \textbf{We introduce a state-aware benchmark for realistic multi-round code review.}
\MCRbench models defect state evolution across review rounds and supports evaluation of cross-round defect identification and state tracking.

    \item \textbf{We develop an automated data construction pipeline for multi-round code review. } 
    The pipeline adopts a state-aware, multi-stage process that supports cross-round defect extraction, merging, and lifecycle tracking.

    \item \textbf{We evaluate the capability boundaries of mainstream LLMs in multi-round code review.} 
    Based on \MCRbench, we assess LLMs' performance on defect identification and lifecycle state prediction, revealing performance variations across interaction depth and defect characteristics.

    \item \textbf{We construct a fine-grained taxonomy of failure root causes.} 
    We employ an open coding process to false positive and false negative cases, resulting in a taxonomy that characterizes the underlying causes of LLM errors in multi-round code review.
\end{itemize}

\section{Related Work}

Earlier code review benchmarks operate on localized code changes at different granularities, including the function/method, diff-hunk, and commit/change levels. At the function/method level, Trans-Review~\cite{tufano2021towards} benchmarks review-driven code transformation by learning typical code revisions from real review activities, while AutoTransform~\cite{thongtanunam2022autotransform} focuses on automated program transformation to support code review, and T5-Review~\cite{tufano2022using} applies pre-trained sequence-to-sequence models to review-oriented code revision tasks. At the diff-hunk level, CodeReviewer~\cite{li2022automating} establishes a large-scale benchmark over isolated code diffs for review tasks such as quality estimation, comment generation and code refinement, while Hybrid-Review-Dataset~\cite{jaoua2025combining} augments diff-hunk–level evaluation with static-analysis signals to support review generation. At the commit/change level, CodeAgent~\cite{tang2024codeagent} evaluates agent-based code review capabilities over commit-level changes, including inconsistency detection and revision suggestion.

With the rapid advancement of LLMs in code understanding and reasoning~\cite{p1,p2,p3,vaswani2017attention,p4,p5,Zhang2026AACRBenchEA,wang2025contextutilization,wang2026reporeasoner}, recent benchmarks have shifted toward evaluating code review at the PR level~\cite{swrbench,codefuse,zhang2026sphinx,hou2024large}, operating on complete PR contexts rather than isolated code changes. SWR-Bench~\cite{swrbench} presents a PR-centric code review benchmark constructed from manually verified GitHub pull requests, providing full project context and structured ground-truth issues to enable objective evaluation of review coverage and quality under realistic review settings. CodeFuse-CR-Bench~\cite{codefuse} introduces a comprehensiveness-aware benchmark for repository-level code review, where each instance supplies rich, multi-faceted PR context—including associated issues, PR metadata, and repository state—to support end-to-end evaluation using both rule-based checks and model-based quality judgments. Sphinx~\cite{zhang2026sphinx} proposes a unified framework for PR review that combines context-rich data generation with a checklist-based evaluation benchmark, enabling structured assessment of review completeness and actionability beyond surface-level similarity metrics.

Despite these advances, most existing benchmarks remain confined to a single-round, static code review paradigm, which fails to capture the iterative nature of real-world code review. In practice, non-trivial code changes typically require multiple cycles of \emph{review $\rightarrow$ revision $\rightarrow$ revalidation} before being merged~\cite{10.1007/s10664-021-10035-z}. This static perspective obscures challenges unique to multi-round settings, particularly defect state tracking (e.g., transitions from \emph{Open} to \emph{Resolved} or \emph{Reopened}) and contextual consistency across review rounds. As a result, existing benchmarks cannot adequately assess LLMs’ capabilities in real-world code review scenarios, emphasizing the urgent need for benchmarks tailored to multi-round code review.

\section{\MCRbench Construction}

\subsection{Benchmark Overview}

\begin{wrapfigure}{r}{0.45\textwidth}
\vspace{-15pt}
    \centering
    \includegraphics[width=1\linewidth]{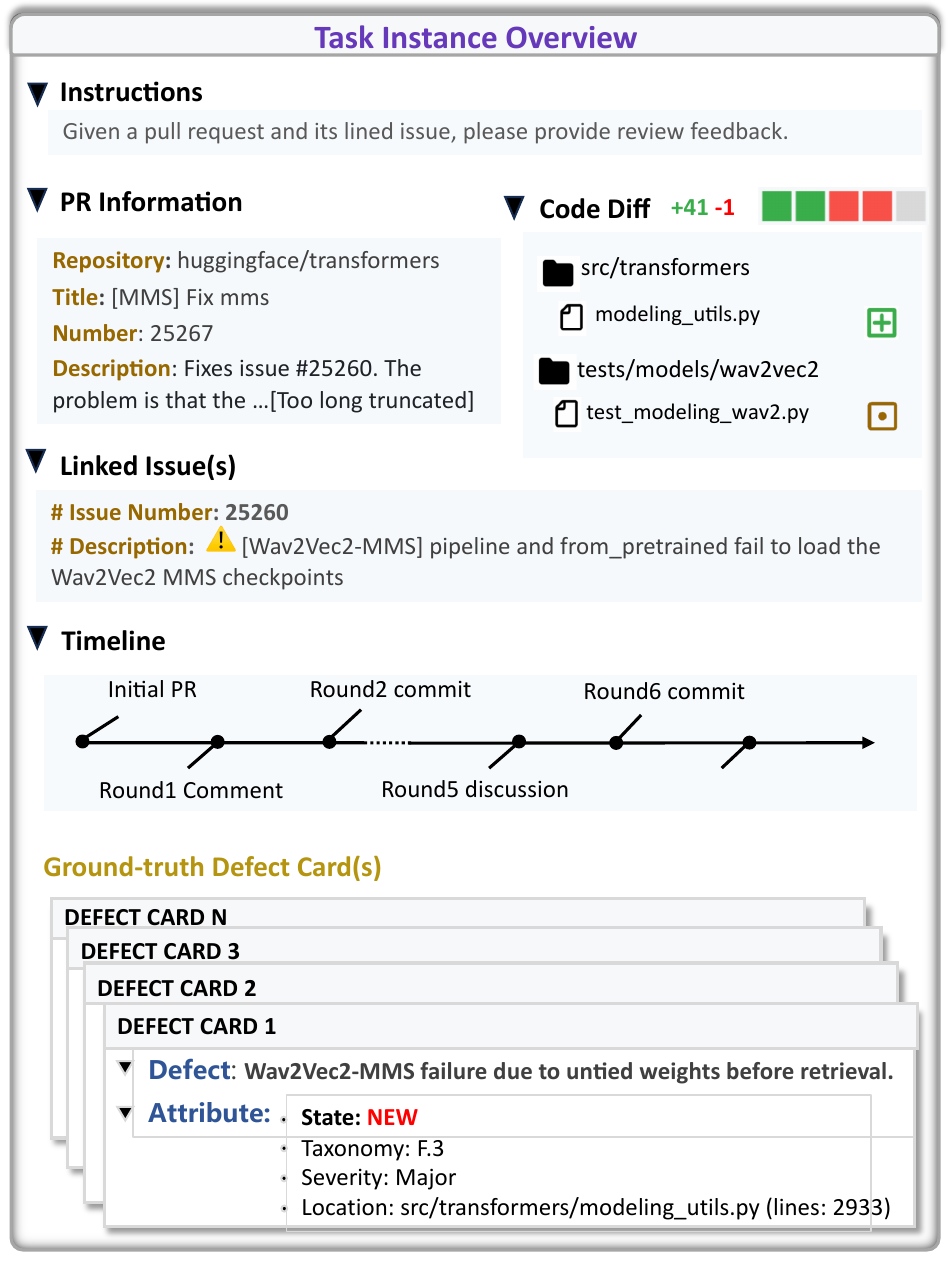}
    \vspace{-25pt}
    \caption{A task in \MCRbench sourced from the GitHub Project huggingface/transformers~\cite{instance}.}
    \label{fig:instance}
    \vspace{-15pt}
\end{wrapfigure}

\MCRbench is a multilingual code review benchmark designed to simulate multi-round review interactions in real-world scenarios. It comprises 2,269 high-quality review tasks spanning the five most popular languages on GitHub. To standardize the evaluation process, we instantiate each review sample as a structured \emph{task instance}. As illustrated in Figure~\ref{fig:instance}, each task consists of two main components: PR related information and the annotated ground truth.

\textbf{PR related information.}
To faithfully reflect LLM performance in multi-round code review, we leverage available PR histories as the basis for constructing evaluation inputs that mirror practical scenarios. Specifically, our input comprises three components.
Task description (\texttt{Instructions}) specifies the code review task to be performed.
Static PR information provides high-level background that remains unchanged across review rounds, including PR metadata (\texttt{PR Information}, e.g., \texttt{Repository}, \texttt{PR Title}, and \texttt{PR Description}) and linked issue records (\texttt{Linked Issue(s)}).
Dynamic review information captures content that evolves across review rounds, including the complete code diff for the current round (\texttt{Code Diff}) and the accumulated review discussions and interactions up to that point (\texttt{Timeline}).

\textbf{Ground truth.}
The evaluation reference for each review round in \MCRbench is represented as a collection of \texttt{Defect Cards}. To facilitate multi-dimensional analysis, the information within each card is organized into three categories: \textbf{(1) Defect Specification}: comprising the natural language \texttt{Defect Description} and precise \texttt{Location} (file paths and line numbers) to uniquely identify the defect; \textbf{(2) Dynamic Lifecycle Status}: recording the defect's \texttt{State} at the current review round (e.g., \emph{New}, \emph{Open}, \emph{Resolved}, or \emph{Reopened}), which serves as the primary ground truth for evaluating state tracking capabilities; and \textbf{(3) Supplementary Metadata}: including \texttt{Defect Taxonomy} and \texttt{Severity}, enabling fine-grained performance analysis across different defect characteristics.

\subsection{Benchmark Construction Pipeline}
\begin{figure}[htbp]
    \centering
    \includegraphics[width=1\textwidth]{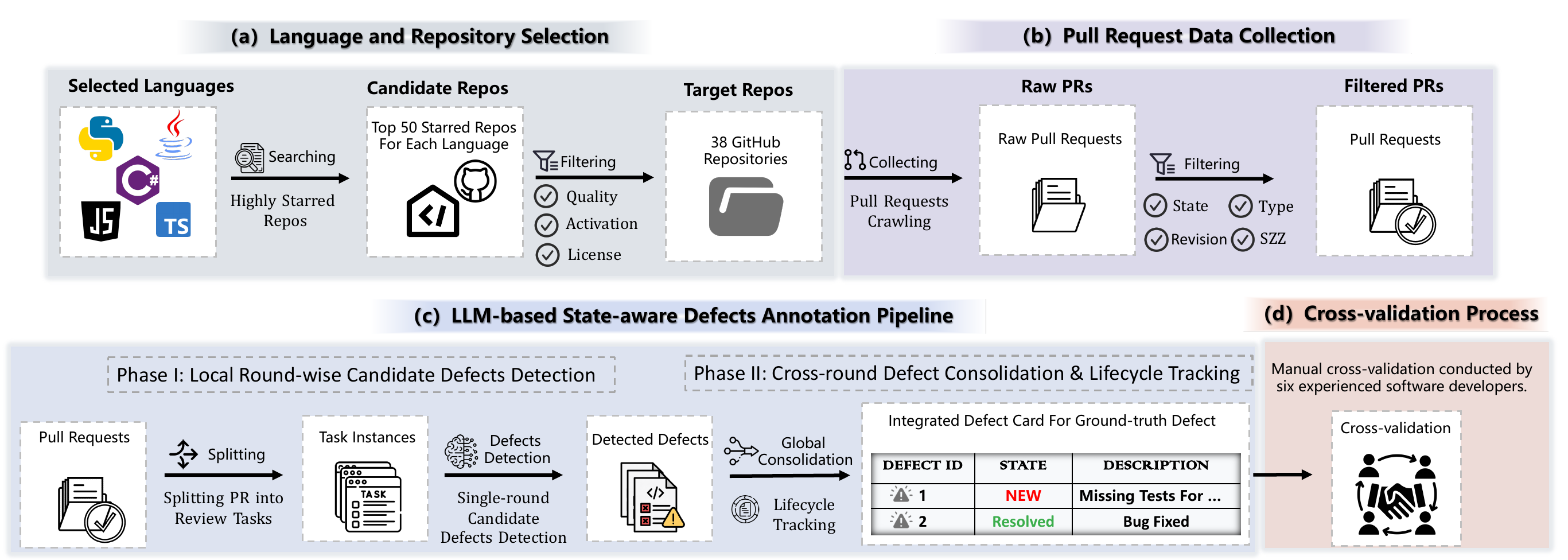}
    \caption{The Construction Pipeline of \MCRbench}
    \label{fig:construction_pipeline}
\end{figure}

As shown in Figure~\ref{fig:construction_pipeline}, \MCRbench's construction pipeline consists of four main steps: (a) language and repository selection, (b) pull request data collection, (c) LLM-based state-aware defect annotation, and (d) the manual cross-validation process.

\subsubsection{Language and Repository Selection.}

To enhance the representativeness and reliability of \MCRbench, we first identify mainstream programming languages and then apply a strict, multi-dimensional repository selection strategy to curate code review data that reflect real-world practice.

\textbf{Programming Language Selection.}
Following the latest GitHub Octoverse report~\cite{Octoverse}, we select the five most active programming languages on GitHub—Python, Java, JavaScript, TypeScript, and C\#. These languages are widely used in modern software development and collectively cover a broad range of application domains, enabling \MCRbench to reflect contemporary engineering practice and support generalizable evaluation~\cite{zheng2024domaincodebench}.

\textbf{Repository Selection.}
Based on the selected programming languages, our repository selection process is designed to maintain the quality and industrial relevance of the source data. To identify well-maintained real-world projects, we employ a set of multi-dimensional filtering criteria for candidate GitHub repositories. Only repositories that satisfy all of the following conditions are retained for subsequent benchmark construction:

\begin{enumerate}[label=(\roman*),leftmargin=2em]

\item Project Maturity and Activity
  \begin{itemize}[leftmargin=5pt]
    \item Community Recognition: Repositories must have more than 100 stars, serving as a coarse indicator of community attention and adoption~\cite{hu2016influence}.
    \item Development Activity: Repositories must exhibit sustained commit activity over the past five years, indicating ongoing development aligned with modern review practices.
    \item Maintenance Activity: Repositories must maintain an issue resolution rate above 40\%, reflecting active maintenance and effective issue handling~\cite{meng2025releaseeval}.
\end{itemize}

\item Collaborative Development
\begin{itemize}[leftmargin=5pt]
    \item Team-Based Development: We retain only repositories with more than 10 contributors to capture realistic team-based collaboration and review dynamics~\cite{meng2025releaseeval}.
    \item PR-Centric Development: Following prior work~\cite{peng2025icodereviewer}, repositories must contain at least 1,500 PRs, indicating sustained PR-based development and sufficient multi-round review data.
\end{itemize}

\item License Legality and Uniqueness
\begin{itemize}[leftmargin=5pt]
    \item License Compliance: Following existing license classification~\cite{Blue,starcoder2}, we retain only repositories under permissive licenses (e.g., MIT) to ensure legal compliance and reproducibility.
    \item Repository Uniqueness: Forked repositories are excluded to avoid duplicated or derivative data, ensuring that review histories reflect original development activity~\cite{swrbench}.
\end{itemize}

\end{enumerate}

\subsubsection{Pull Request Data Collection.}

Based on the filtered repositories, we further process all PRs within these repositories to obtain high-quality samples for multi-round code review evaluation. To this end, we design and implement a filtering pipeline over the raw PR data (as illustrated in Figure~\ref{fig:construction_pipeline}(b)), which progressively distills high-quality multi-round review instances.
\blackcircled{1} \textbf{State Filtering.} We retain only PRs with the \emph{Merged} status, as merged PRs indicate that the code changes have been reviewed and accepted by core maintainers and integrated into the project, providing a basic guarantee of solution correctness.
\blackcircled{2} \textbf{Type and Scale Filtering.} We retain PRs that modify meaningful code logic while respecting model context constraints. Specifically, we remove PRs that only affect non-code files (e.g., documentation, images) and exclude PRs with more than 10 initial commits, as such changes are likely to exceed model context limits. 
\blackcircled{3} \textbf{Multi-Round Interaction.} We enforce a strict multi-round interaction constraint, which is central to \MCRbench. To capture genuine dynamic review processes, we require retained PRs to exhibit an iterative \textbf{\emph{< commit $\rightarrow$ discussion $\rightarrow$ revision >}} loop. Specifically, authors must submit new commits in direct response to reviewer feedback, rather than leaving comments without code changes. Additionally, we filter out noise from automated bots to ensure all interactions reflect human-to-human collaboration.
\blackcircled{4} \textbf{Quality Control.} Inspired by prior work~\cite{swrbench}, we apply SZZ-based quality control to detect latent defects in merged PRs. Since merging does not necessarily guarantee defect-free code due to potential human oversight, we use the SZZ-2 algorithm~\cite{szz1,szz2} to retrospectively analyze whether a PR is later identified as bug-inducing during subsequent development. PRs that are confirmed to have introduced new bugs after merging are excluded from the dataset.

 \subsubsection{LLM-based State-aware Defects Annotation Pipeline.}

To enable scalable and consistent state-aware annotation for multi-round code review, we design an automated pipeline (Figure~\ref{fig:construction_pipeline}(c))~\cite{yang2025benchmarkbuilders}. The pipeline takes complete PR history, including round-level diffs, reviewer comments, and subsequent developer revisions, as input and produces unified ground-truth \texttt{Defect Cards} with lifecycle states across review rounds. It follows a divide-and-conquer strategy of \emph{``local detection first, global tracking later''}: it first extracts round-specific defect candidates from local review contexts, and then consolidates them into cross-round defect annotations. This strategy avoids directly annotating long and iterative review histories in a single pass, where round boundaries and cross-round defect identities can become difficult to distinguish, thereby reducing annotation reliability.

\textbf{Phase I: Local Round-wise Candidate Defects Detection.} The primary goal of Phase I is to maximize recall by improving the coverage of potential defects at the round level. To mitigate attention degradation caused by long contexts, we decompose the entire PR lifecycle into a sequence of independent review task instances. For the $t$-th review round, the LLM is provided only with the code diffs and review comments from this round, and is instructed to identify all reviewer-raised defects, producing a set of round-specific candidate defects. The outputs of this stage are intermediate defect candidates rather than final benchmark annotations: each candidate is treated as a round-local and static defect instance, without cross-round identity linkage or lifecycle state modeling.

\textbf{Phase II: Cross-round Defect Consolidation and Lifecycle Tracking}. Phase II transforms the local candidates from Phase I into state-aware benchmark annotations at the PR level. We first aggregate all candidate defects identified in Phase I into a global pool, and then leverage the semantic reasoning capabilities of LLMs to identify and merge entries that refer to the same underlying defect across different review rounds. This step yields a set of unified defect representations spanning the full review process. Based on these unified defects, the model further reasons over subsequent author commits, historical reviewer feedback, and prior defect mentions to infer the lifecycle state of each defect at the end of every review round. In this way, the final annotation is not merely a collection of isolated round-wise comments, but a coherent sequence of defect cards that explicitly captures cross-round persistence, resolution, and reopening.
In real-world code review practice, defects typically evolve through the following lifecycle states:

\begin{itemize}[leftmargin=10pt]
\item \textbf{New}: the defect is raised for the first time in the current review round.
\item \textbf{Open}: the defect has been identified but remains unresolved up to the current review round.
\item \textbf{Resolved}: the defect is addressed by code changes and verified within the current review round.
\item \textbf{Reopened}: the defect was resolved in a previous round but reappears in a subsequent round.
\end{itemize}

\textbf{Consistency Filtering.} To improve the scalability and reliability of the LLM-based annotation pipeline, we introduce an additional consistency filtering step, following common reliability-oriented practices in prior work~\cite{swrbench,ahmed2025annotation,llift2024}. Specifically, we independently execute the annotation pipeline three times for each task. A task is retained only if the generated defect cards are fully consistent across all runs in terms of both lifecycle state transitions and semantic defect descriptions. Any sample that exhibits discrepancies across runs, such as conflicting state predictions (e.g., \emph{Resolved} versus \emph{Open}), is treated as an uncertain case and discarded.

\subsubsection{Cross-validation Process.}

After consistency filtering, we conduct manual cross-validation to verify the retained high-confidence defect cards. Six developers with more than five years of programming experience form the annotator pool and re-examine each retained multi-round code review task.
Specifically, each retained task instance is first assigned to two primary annotators for independent verification. They separately inspect the generated defect cards and assess whether the defect descriptions are accurate, whether the cross-round consolidation is rational, and whether the lifecycle states correctly align with the corresponding code changes and review discussions. Only annotations on which the two primary annotators agree are directly accepted. We measure the agreement between the two primary annotators using Cohen's kappa~\cite{cohen1960coefficient}, obtaining a score of 0.87, which indicates high agreement.
When disagreements arise between the two primary annotators, the task is escalated to a third annotator who serves as the arbitrator. The three annotators then jointly examine the complete PR history, code changes, and review discussions, and resolve discrepancies through discussion until a final consensus is reached. Only after this consensus-based verification is the task accepted into the benchmark dataset.

\section{\MCRbench}

\subsection{Statistics of \MCRbench}
\begin{figure}[htbp]
    \centering
    \includegraphics[width=0.9\textwidth]{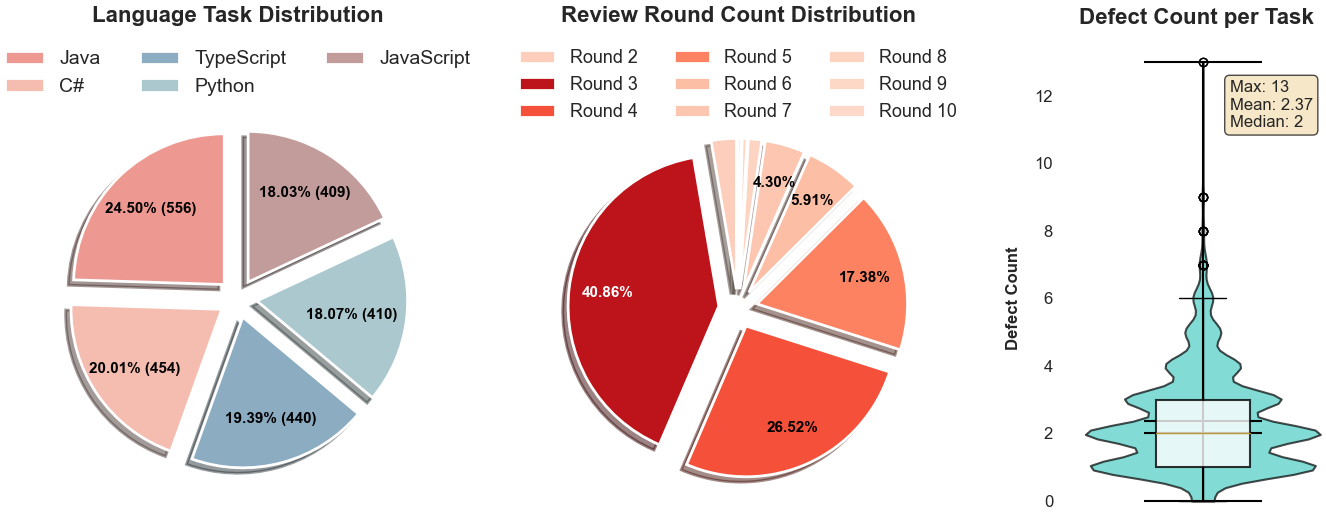}
    \caption{Statistics of \MCRbench: language distribution, review round distribution, and defect number statistics.}
    \label{fig:statistics}
\end{figure}

To profile the composition of \MCRbench, we perform a multi-dimensional statistical analysis on its 2,269 manually validated task instances. The analysis covers five aspects: programming language distribution, interaction rounds distribution, defect statistics, defect severity, and defect taxonomy.

\textbf{Language Distribution.}
As shown in Figure~\ref{fig:statistics}(a), \MCRbench spans five mainstream programming languages with a relatively even distribution. Java accounts for 556 instances (24.50\%), followed by C\# (454, 20.01\%) and TypeScript (440, 19.39\%), while Python and JavaScript contribute comparable shares with 410 (18.07\%) and 409 (18.03\%) instances, respectively.

\textbf{Interaction Round Distribution.}
Figure~\ref{fig:statistics}(b) illustrates the distribution of interaction rounds across tasks. All tasks contain at least two review rounds. Among them, tasks with three rounds (Round~3) constitute the largest proportion at 40.86\%, followed by four-round (Round~4) and five-round (Round~5) tasks, which account for 26.52\% and 17.38\%, respectively. Notably, the majority of tasks involve three or more rounds of interaction, indicating that most samples in \MCRbench require models to handle long-range multi-round interactions rather than simple single-round responses.

\textbf{Statistics of Number of Defects Per Task.}
The violin plot in Figure~\ref{fig:statistics}(c) visualizes the distribution of the number of defects per task. Across the entire dataset, the mean number of defects per task is 2.37, the median is 2, and the maximum reaches 13 defects in a single task. Most tasks contain between one and four defects. This moderate defect density ensures sufficient challenge for evaluating model recall while avoiding excessive contextual complexity that could obscure review signals.

\begin{table}[htbp]
\caption{Defect taxonomy and distribution in \MCRbench.}
\label{tab:defect_taxonomy}
\centering
\scriptsize
\setlength{\tabcolsep}{6pt}
\renewcommand{\arraystretch}{1.15}
\begin{tabular}{l p{2.4cm} p{7.2cm} r}
\toprule
ID & Name & Description & Ratio \\
\midrule
E & Evolvability & Improvements that enhance future maintainability without changing externally observable behavior. & -- \\

\quad E.1 & Documentation & Changes to in-code information to improve human understanding. & -- \\
\quad\quad E.1.1 & Textual & Comments, identifier or name changes, README text, and error-message wording. & 19.19\% \\
\quad\quad E.1.2 & Language-Supported & Docstrings, type annotations, and language-supported documentation. & 5.56\% \\

\quad E.2 & Visual Representation & Layout or style changes for better readability. & 10.25\% \\

\quad E.3 & Structure & Changes to code organization or architecture. & -- \\
\quad\quad E.3.1 & Organization & Moving, splitting, merging, or extracting code. & 9.61\% \\
\quad\quad E.3.2 & Solution Approach & Refactoring without altering externally observable semantics. & 11.79\% \\

\quad E.4 & Verification / Test & Requests to add or improve test cases. & 8.98\% \\

\midrule
F & Functional & Changes that alter or fix software behavior or runtime properties. & -- \\

\quad F.1 & Interface & API contracts and interactions between components. & 5.19\% \\
\quad F.2 & Logic & Algorithms, business rules, and control-flow correctness. & 13.44\% \\
\quad F.3 & Resource & Management of variables, resources, and state. & 3.64\% \\
\quad F.4 & Check & Guards for unhandled states and error conditions. & 4.82\% \\
\quad F.5 & Support & Interactions with external systems or dependencies. & 3.16\% \\
\quad F.6 & Performance & Efficiency and performance-related concerns. & 3.74\% \\
\quad F.7 & Security \& Privacy & Security and privacy risks. & 0.63\% \\

\bottomrule
\end{tabular}
\end{table}

\textbf{Statistics of Defect Type Taxonomy.}
To enable a more in-depth analysis of LLM performance across different defect types, we adopt a fine-grained taxonomy to categorize all annotated defects in \MCRbench. As shown in Table~\ref{tab:defect_taxonomy}, defects are organized into two major groups: \emph{Functional} defects, which relate to program behavior and correctness, and \emph{Evolvability} defects, which concern maintainability and long-term code quality.
The taxonomy is adapted from established classification frameworks in prior work~\cite{swrbench,10.1145/2597073.2597082,10.1007/s10664-022-10205-7}, and we further extend it with test- and security-related categories to improve granularity and better align with the broader software engineering literature.

\begin{itemize}[leftmargin=10pt]
\item \textbf{E.4 Verification / Test:} This category covers review comments requesting the addition or improvement of test cases, reflecting testing as a distinct concern in code review~\cite{E41, E42}.
\item \textbf{F.7 Security \& Privacy:} This category covers security vulnerabilities and privacy risks, distinguishing them from other functional defects in line with prior empirical studies~\cite{F71, F72, F73}.
\end{itemize}

As shown Table~\ref{tab:defect_taxonomy}, overall, the distribution of defects in \MCRbench broadly aligns with prior code review datasets~\cite{swrbench,10.1145/2597073.2597082,10.1007/s10664-022-10205-7}. \emph{Functional defects} mainly involve logic-related defects, followed by interface-related problems and missing or insufficient checks, indicating that many reviews focus on substantive correctness and robustness concerns. \emph{Evolvability defects} primarily address code readability, structure, and maintainability, with textual defects such as naming and comments appearing frequently, alongside refactoring-oriented concerns related to solution structure and design. These defect types capture the dual role of code review in identifying correctness defects and improving long-term code quality and maintainability.

\textbf{Statistics of Defect Severity.}
To examine the sensitivity of LLMs to defects of different severity levels during code review, we annotate all defects in \MCRbench using a standard severity classification scheme introduced in prior work~\cite{10.1145/3194095.3194101}.

As shown in Table~\ref{tab:severity_levels}, the distribution of defect severity in \MCRbench broadly aligns with observations from real-world code review practice, where review feedback predominantly targets lower-risk defects, while higher-severity problems are identified less often but remain important~\cite{10.1145/2597073.2597082,10.1007/s10664-023-10411-x}. Specifically, defects labeled as \emph{Trivial}, \emph{Minor}, and \emph{Normal} are commonly observed in \MCRbench and primarily relate to routine functional adjustments and code quality improvements. In addition, \MCRbench includes a range of higher-severity defects. \emph{Major} defects capture problems with notable impact on core functionality that are frequently identified during multi-round code review, and although \emph{Blocker} and \emph{Critical} defects occur less often, they correspond to cases that can hinder code integration or pose risks to system stability. Overall, this distribution covers both commonly encountered, lower-risk defects and higher-severity defects with substantial impact, enabling evaluation across a broad range of risk levels.

\begin{table}[htbp]
\caption{Defect severity levels and their distribution in \MCRbench.}
\label{tab:severity_levels}
\centering
\scriptsize
\setlength{\tabcolsep}{6pt}
\renewcommand{\arraystretch}{1.15}
\begin{tabular}{l p{10.6cm} r}
\toprule
\textbf{Severity} & \textbf{Description} & \textbf{Ratio} \\
\midrule
Trivial & Cosmetic-only problems, such as misspellings or misaligned text. & 22.26\% \\

Minor & Causes a minor loss of functionality, or a defect for which an easy workaround is available. & 24.54\% \\

Normal & A regular defect that leads to partial loss of functionality under specific conditions. & 29.73\% \\

Major & Causes a major loss of functionality. & 20.21\% \\

Blocker & Blocks development and/or testing work; the change cannot proceed or be merged until the defect is fixed. & 2.32\% \\

Critical & Causes crashes, data loss, or severe memory leaks. & 0.93\% \\

\bottomrule
\end{tabular}
\end{table}

\subsection{Characteristics of \MCRbench}

\textbf{(1) Multi-round, State-aware Code Review Benchmark.}
To the best of our knowledge, \MCRbench is the first benchmark that formulates code review as a dynamic, multi-round interaction process at the PR level. As shown in Table~\ref{tab:benchmark_comparison}, unlike existing benchmarks that primarily model review as a single-round task, \MCRbench captures the iterative nature of real-world code review, with each task containing at least two review rounds and an average of 3.8 rounds. Beyond merely defect identification, \MCRbench continuously tracks defect state transitions across review rounds, distinguishing newly introduced, open, resolved, and reappearing defects, and thus requires LLMs to reason consistently over multi-round contexts.

\textbf{(2) Rigorous Construction Pipeline with Multi-stage Verification.}
Unlike existing benchmarks that are typically built from code diffs, \MCRbench establishes a rigorous, multi-stage construction framework tailored to multi-round review scenarios, especially our LLM-based defect annotation pipeline. This pipeline follows a \emph{``local-first, global-later''} strategy, where candidate defects are first extracted from individual review rounds and then consolidated at the PR level through cross-round defect merging and state tracking. To obtain reliable annotations, we independently execute our two-stage LLM-based annotation pipeline three times per task and retain only tasks whose resulting defect cards are fully consistent across runs, following a reliability-oriented quality-control design commonly adopted in prior work~\cite{swrbench,ahmed2025annotation,llift2024}. Finally, all tasks undergo manual cross-validation as a final verification step to resolve potential conflicts, further improving the stability and accuracy of defect annotations across evolving contexts.

\textbf{(3) Multi-language Benchmark with Structured Annotations and Rich Context and.}
\MCRbench is a multi-round code review benchmark that enables fine-grained analysis of LLM performance across diverse defect characteristics. It covers five widely used programming languages: Python, Java, JavaScript, TypeScript, and C\#. Each defect instance in \MCRbench is annotated with structured defect categories, severity levels, and round-specific lifecycle states, enabling detailed analysis of defect types, severity, and state evolution. Moreover, \MCRbench operates at the PR level with complete repository context, including linked issues and historical review interactions, supporting realistic and context-rich evaluation of code review models.

\definecolor{tableblue}{RGB}{210,230,245}
\begin{table}[t]
\centering
\scriptsize
\setlength{\tabcolsep}{4pt}
\renewcommand{\arraystretch}{1.15}
\caption{Comparison between popular code review benchmarks and \MCRbench.}
\label{tab:benchmark_comparison}
\vspace{-10pt}
\begin{tabular}{l c r c l r c >{\centering\arraybackslash}m{100pt}}
\toprule
\multicolumn{1}{c}{\multirow{2}{*}{\textbf{Dataset}}} &
\multicolumn{1}{c}{\multirow{2}{*}{\makecell{\textbf{Review}\\\textbf{Granularity}}}} &
\multicolumn{3}{c}{\textbf{\#Review Rounds}} &
\multicolumn{1}{c}{\multirow{2}{*}{\textbf{\#Tasks}}} &
\multicolumn{1}{c}{\multirow{2}{*}{\textbf{\#Languages}}} &
\multirow{2}{*}{\textbf{Context Scope}} \\
\cline{3-5}
& & \textbf{Min} & \textbf{Avg} & \textbf{Max} & & & \\
\midrule

Trans-Review~\cite{tufano2021towards}        & Method        & 1 & 1.0 & 1 & 1,719  & 1 & None \\
AutoTransform~\cite{thongtanunam2022autotransform}       & Method        & 1 & 1.0 & 1 & 14,750 & 1 & None \\
T5-Review~\cite{tufano2022using}           & Method        & 1 & 1.0 & 1 & 17,194 & 1 & None \\
CodeReviewer~\cite{li2022automating}        & Diff Hunk     & 1 & 1.0 & 1 & 10,000 & 9 & None \\
CR-Agent-Dataset~\cite{tang2024codeagent}    & Commit        & 1 & 1.0 & 1 & 3,545  & 9 & Related Source Code \\
Hybrid-Review-Dataset~\cite{jaoua2025combining}  & Diff Hunk     & 1 & 1.0 & 1 & 1,245  & 1 & Related Source Code \\
\midrule
SWR-Bench~\cite{swrbench}           & Pull Request  & 1 & 1.0 & 1 & 1,000  & 1 & Complete Codebase \\
CodeFuse-CR-Bench~\cite{codefuse}   & Pull Request  & 1 & 1.0 & 1 & 601    & 1 & Complete Codebase + Linked Issues \\

\midrule
\rowcolor{tableblue}
\textbf{\MCRbench} & \textbf{Pull Request} & \textbf{2} & \textbf{3.8} & \textbf{10} & \textbf{2,269} & \textbf{5} &
\textbf{Complete Codebase + Linked Issues + History Actions} \\
\bottomrule
\end{tabular}
\end{table}

\section{Experimental Design}

\subsection{Research Questions}
\label{sec:RQ}

Our experiments intend to answer the following research questions (RQs):

\begin{itemize}[leftmargin=10pt]
    \item \textbf{RQ1 (General Performance): How do mainstream LLMs perform on \MCRbench?}

    \item \textbf{RQ2 (Performance Evolution across Rounds): How does LLM performance evolve as the review process progresses?}

    \item \textbf{RQ3 (Root Cause Analysis): What are the primary factors contributing to model failures (false positives and false negatives) in multi-round code review?}

    \item \textbf{RQ4 (Comment Quality): What is the quality of review comments generated by pure LLM prompting and ACR baselines under practitioner-oriented criteria?}
    
\end{itemize}

\subsection{Studied Models and Baselines}
\label{sec:LLM}

\textbf{Studied LLMs.}
To evaluate LLM performance in multi-round code review scenarios, we study a diverse set of representative models drawn from both closed-source commercial systems and open-source model families. The selected models span diverse training paradigms and context modeling characteristics. Our evaluation includes three widely used commercial models—GPT-5.2, Claude-Haiku-4.5, and Gemini-3-Flash~\cite{gpt,anthropic,gemini}—as well as four open-source models—DeepSeek-V3.2, Qwen3-Max, GLM-4.7, and Kimi-k2~\cite{deepseek,glm,kimi,qwen}. This selection enables a balanced evaluation across industrial and open-source model ecosystems and provides a broad view of current LLM capabilities in multi-round code review.

\textbf{Studied ACR Baselines.}
We evaluate two representative automated code review (ACR) baselines, PR-Agent and Hybrid-Review, which correspond to two common ACR designs: an agent-based PR review pipeline and a static-analysis-assisted review generation pipeline. Existing PR-level ACR methods are mostly designed for single-round review settings, and thus cannot be directly applied to the multi-round, state-aware setting of \MCRbench. We therefore select these two representative baselines and make minimal necessary adaptations so that they can operate under the evaluation protocol of \MCRbench.

\begin{itemize}[leftmargin=10pt]
    \item PR-Agent~\cite{pragent} is an agent-based PR review baseline. Given a PR, it analyzes the code diff and related PR context, and then generates review feedback through an agent-based review workflow.
    \item Hybrid-Review~\cite{jaoua2025combining} is a static-analysis-assisted review generation baseline. It first collects static-analysis warnings from the changed code and then uses these warnings as defect-related signals to guide LLM-generated review comments.
\end{itemize}

\vspace{-10pt}
\subsection{Evaluation Metric}
\label{sec:Metric}

The unstructured nature of model-generated code review comments poses inherent evaluation challenges, necessitating an empirical assessment of how well existing automatic metrics align with human judgments. To this end, we survey commonly used evaluation metrics and conduct a pre-study to analyze their agreement with human annotations, based on which we select the most suitable metric for subsequent experiments.

\textbf{Candidate Metrics.}
We investigate a diverse set of candidate metrics commonly used in prior work, ranging from traditional lexical overlap metrics to LLM-based semantic and functional evaluation approaches. Lexical overlap metrics such as BLEU-4~\cite{Blue} and ROUGE-L~\cite{rouge} are widely used in existing code review benchmarks for their simplicity~\cite{li2022automating,meng2025code,li2025issue,s23052551,10.1007/s10515-025-00491-y}, but they rely on surface-level lexical or structural similarity. This reliance limits their ability to capture the semantic correctness of review comments, particularly in multi-round settings where similar defects may be expressed in varied ways or implied through context, leading to underestimation of valid defect-identifying comments.

To address these limitations, recent studies have proposed LLM-based evaluation metrics that leverage the reasoning and semantic understanding capabilities of LLMs. Broadly, these metrics fall into two primary paradigms: \emph{LLM Scoring}~\cite{codefuse} and \emph{LLM-Hit-Judge}~\cite{swrbench}. \emph{LLM Scoring} assigns a quality score to a generated review comment by prompting a judge LLM with the code context and the model output, where the score reflects the judge’s assessment of usefulness, relevance, and clarity. In contrast, \emph{LLM-Hit-Judge} focuses on defect-level correctness by verifying whether a generated comment successfully identifies a specific ground-truth defect through binary judgment. In our pre-study, we employ three models as judge LLMs—\emph{GPT-5.2-pro}, \emph{Claude-Opus-4.5}, and \emph{Gemini-3-pro}—to examine the robustness and representativeness of these LLM-based evaluation metrics.

\textbf{Pre-study Design and Human Annotation.}
To identify the evaluation metric that is most consistent with human judgment, we randomly sample 10\% of the generated review comments produced by the selected LLMs on \MCRbench as evaluation instances. We invite four senior developers, each with more than five years of programming experience, to serve as human annotators and perform cross-validation on the sampled comments. For each instance, the annotators label the true defect hit rate (\emph{Human Hit Rate}), which serves as the ground truth for subsequent analysis.

\textbf{Agreement Measurement.}
We then apply all candidate automatic evaluation metrics to the same set of review comments. For lexical overlap metrics, scores are computed by comparing the generated comments with the corresponding reference annotations. For LLM-based metrics, in our pre-study, we employ three strong models as judge LLMs (\emph{GPT-5.2-pro}, \emph{Claude-Opus-4.5}, and \emph{Gemini-3-pro}).
The resulting metric scores are then correlated with the Human Hit Rate to assess their consistency with human judgments.

We use the \emph{Quadratic Weighted Kappa} (QWK)~\cite{QWK1,QWK2,QWK3} to quantify the agreement between the scores produced by automatic evaluation metrics and the Human Hit Rate. QWK assigns lower agreement scores when the discrepancy between human judgments and automatic metric scores is larger, thereby reflecting how closely an automatic metric aligns with human evaluation.

\begin{wraptable}{r}{0.45\textwidth}
\vspace{-10pt}
\centering
\footnotesize
\setlength{\tabcolsep}{10pt}
\renewcommand{\arraystretch}{1}
\caption{Agreement between automatic metrics and Human Hit Rate measured by QWK on the sampled \MCRbench evaluation set.}
\label{tab:qwk_agreement}
\vspace{-10pt}
\begin{tabular}{lc}
\toprule
\textbf{Metric} & \textbf{QWK} \\
\midrule
ROUGE-L & 0.21 \\
BLEU-4 & 0.27 \\
\midrule
LLM Scoring (Claude Opus 4.5) & 0.58 \\
LLM Scoring (Gemini 3 Pro) & 0.55 \\
LLM Scoring (GPT-5.2 Pro) & 0.60 \\
\midrule
LLM-Hit-Judge (Claude Opus 4.5) & 0.70 \\
LLM-Hit-Judge (Gemini 3 Pro) & 0.67 \\
\textbf{LLM-Hit-Judge (GPT-5.2 Pro)} & \textbf{0.73} \\

\bottomrule
\end{tabular}
\vspace{-10pt}
\end{wraptable}
\textbf{Pre-study Results.}
Table~\ref{tab:qwk_agreement} reports the agreement between different automatic evaluation metrics and the Human Hit Rate, measured by QWK. Traditional text-overlap-based metrics, such as ROUGE-L and BLEU-4, achieve low QWK scores of 0.21 and 0.27, indicating limited agreement with human judgments and a limited ability to reflect human-annotated defect coverage at the hit-rate level. In contrast, LLM-based evaluation metrics show substantially higher agreement with human judgments. LLM Scoring achieves QWK values ranging from 0.55 to 0.60 across different judge models, corresponding to moderate agreement, while the defect-coverage-oriented LLM-Hit-Judge attains the highest QWK scores across all three judge models, with a peak value of 0.73. Based on these results, we adopt LLM-Hit-Judge with \emph{GPT-5.2-pro} as the judge model as the primary evaluation metric in all subsequent experiments.
For completeness, we additionally report the raw scores of all studied metrics in our public repository.\footnote{\url{https://github.com/DeepSoftwareAnalytics/MCR-bench}}

\section{Experimental Results}

\vspace{-15pt}
\begin{table}[htbp]
\centering
\tiny
\setlength{\tabcolsep}{3pt}
\renewcommand{\arraystretch}{1.05}
\caption{Performance across programming languages and overall results (Precision / Recall / F1).}
\label{tab:overall_performance}
\vspace{-10pt}
\begin{tabular}{l rcl rcl rcl rcl rcl rcl}
\toprule
\multirow{2}{*}{\textbf{Model}}
& \multicolumn{3}{c}{\textbf{Python}}
& \multicolumn{3}{c}{\textbf{Java}}
& \multicolumn{3}{c}{\textbf{JavaScript}}
& \multicolumn{3}{c}{\textbf{TypeScript}}
& \multicolumn{3}{c}{\textbf{C\#}}
& \multicolumn{3}{c}{\textbf{Overall}} \\
\cmidrule(lr){2-4}
\cmidrule(lr){5-7}
\cmidrule(lr){8-10}
\cmidrule(lr){11-13}
\cmidrule(lr){14-16}
\cmidrule(lr){17-19}
& \textbf{Prec.} & \textbf{Rec.} & \textbf{F1}
& \textbf{Prec.} & \textbf{Rec.} & \textbf{F1}
& \textbf{Prec.} & \textbf{Rec.} & \textbf{F1}
& \textbf{Prec.} & \textbf{Rec.} & \textbf{F1}
& \textbf{Prec.} & \textbf{Rec.} & \textbf{F1}
& \textbf{Prec.} & \textbf{Rec.} & \textbf{F1} \\
\midrule
Claude Haiku 4.5 & 0.556 & 0.619 & 0.540 & 0.561 & 0.643 & 0.548 & 0.640 & 0.664 & 0.599 & 0.552 & 0.594 & 0.520 & 0.587 & 0.631 & 0.548 & 0.579 & 0.630 & 0.551 \\
DeepSeek V3.2 & 0.620 & 0.442 & 0.484 & 0.605 & 0.481 & 0.495 & 0.673 & 0.476 & 0.519 & 0.590 & 0.430 & 0.460 & 0.638 & 0.447 & 0.491 & 0.625 & 0.455 & 0.490 \\
Gemini 3 Flash & 0.649 & 0.548 & 0.554 & 0.594 & 0.562 & 0.529 & 0.653 & 0.564 & 0.553 & 0.571 & 0.491 & 0.486 & 0.638 & 0.540 & 0.538 & 0.621 & 0.541 & 0.532 \\
GLM-4.7 & 0.597 & 0.446 & 0.476 & 0.604 & 0.465 & 0.491 & 0.626 & 0.464 & 0.489 & 0.553 & 0.402 & 0.425 & 0.602 & 0.432 & 0.467 & 0.596 & 0.442 & 0.470 \\
GPT-5.2 & 0.651 & 0.563 & 0.563 & 0.582 & 0.574 & 0.527 & 0.682 & 0.566 & 0.573 & 0.584 & 0.545 & 0.516 & 0.609 & 0.557 & 0.533 & 0.622 & 0.561 & 0.542 \\
Kimi K2 & 0.520 & 0.357 & 0.388 & 0.453 & 0.352 & 0.364 & 0.514 & 0.365 & 0.391 & 0.443 & 0.314 & 0.331 & 0.522 & 0.356 & 0.389 & 0.490 & 0.349 & 0.373 \\
Qwen3 Max & 0.539 & 0.386 & 0.411 & 0.507 & 0.412 & 0.414 & 0.544 & 0.401 & 0.422 & 0.506 & 0.367 & 0.388 & 0.199 & 0.140 & 0.153 & 0.459 & 0.341 & 0.357 \\
\bottomrule
\end{tabular}
\vspace{-20pt}
\end{table}

\subsection{General Performance (RQ1)}

This subsection investigates the overall performance of mainstream LLMs on \MCRbench in multi-round code review scenarios. To thoroughly assess their capability in handling multi-round code review tasks, we structure our evaluation around two core dimensions: (i) the ability of LLMs to identify defects under multi-round code review settings, and (ii) their ability to track defect states across successive review rounds. Based on these two dimensions, we further conduct in-depth analyses to examine model behavior under different defect characteristics and to reveal common strengths and limitations in multi-round code review settings.

\subsubsection{Defect Detection Performance}
Table~\ref{tab:overall_performance} presents the performance of all studied LLMs on defect identification across five programming languages. Overall, mainstream LLMs exhibit only modest performance on \MCRbench, with F1 scores largely in the low-to-medium range and the best-performing model slightly exceeding 0.55, indicating persistent challenges in defect identification and contextual reasoning under multi-round code review settings.

From a model-centric perspective, noticeable performance differences are observed across different LLMs. Claude Haiku~4.5 and GPT-5.2 consistently achieve higher F1 scores across most programming languages, with relatively balanced precision and recall, reflecting more stable defect identification behavior. In contrast, LLMs such as Qwen3-Max and Kimi-K2 tend to make conservative predictions in multi-round code review, which leads to consistently low recall and, consequently, lower overall F1 scores.

From a language-centric perspective, performance differences are observed across programming languages. Models generally achieve higher F1 scores on Python and JavaScript, where relative performance rankings remain more stable, compared to Java, TypeScript, and C\#. For languages such as TypeScript and C\#, most models experience a pronounced drop in recall, and some perform substantially worse than their overall averages.

\begin{table*}[t]

\centering
\caption{Performance of baselines across programming languages and overall results (Precision / Recall / F1).}
\vspace{-10pt}
\label{tab:acr_tool_results}
\tiny
\setlength{\tabcolsep}{2.1pt}
\renewcommand{\arraystretch}{1.2}
\begin{tabular}{ll rcl rcl rcl rcl rcl rcl}
\toprule
\multirow{2}{*}{\textbf{Baselines}} & \multirow{2}{*}{\textbf{Model}}
& \multicolumn{3}{c}{\textbf{Python}}
& \multicolumn{3}{c}{\textbf{Java}}
& \multicolumn{3}{c}{\textbf{JavaScript}}
& \multicolumn{3}{c}{\textbf{TypeScript}}
& \multicolumn{3}{c}{\textbf{C\#}}
& \multicolumn{3}{c}{\textbf{Overall}} \\
\cmidrule(lr){3-5}
\cmidrule(lr){6-8}
\cmidrule(lr){9-11}
\cmidrule(lr){12-14}
\cmidrule(lr){15-17}
\cmidrule(lr){18-20}
& & \textbf{Prec.} & \textbf{Rec.} & \textbf{F1} & \textbf{Prec.} & \textbf{Rec.} & \textbf{F1} & \textbf{Prec.} & \textbf{Rec.} & \textbf{F1} & \textbf{Prec.} & \textbf{Rec.} & \textbf{F1} & \textbf{Prec.} & \textbf{Rec.} & \textbf{F1} & \textbf{Prec.} & \textbf{Rec.} & \textbf{F1} \\
\midrule
\multirow{6}{*}{PR-Agent}
& Claude Haiku 4.5 & 0.273 & 0.236 & 0.232 & 0.299 & 0.306 & 0.276 & 0.311 & 0.278 & 0.268 & 0.290 & 0.260 & 0.248 & 0.308 & 0.266 & 0.259 & 0.296 & 0.269 & 0.257 \\
& DeepSeek V3.2 & 0.462 & 0.389 & 0.391 & 0.508 & 0.527 & 0.466 & 0.474 & 0.422 & 0.407 & 0.463 & 0.421 & 0.403 & 0.489 & 0.423 & 0.414 & 0.479 & 0.436 & 0.416 \\
& Gemini 3 Flash & 0.271 & 0.127 & 0.161 & 0.315 & 0.148 & 0.188 & 0.351 & 0.181 & 0.222 & 0.292 & 0.147 & 0.182 & 0.328 & 0.152 & 0.193 & 0.311 & 0.151 & 0.189 \\
& GPT-5.2 & 0.352 & 0.283 & 0.279 & 0.393 & 0.420 & 0.356 & 0.389 & 0.353 & 0.323 & 0.340 & 0.325 & 0.289 & 0.383 & 0.318 & 0.303 & 0.371 & 0.340 & 0.310 \\
& Kimi K2 & 0.381 & 0.283 & 0.301 & 0.488 & 0.411 & 0.410 & 0.456 & 0.345 & 0.364 & 0.413 & 0.310 & 0.325 & 0.436 & 0.313 & 0.333 & 0.435 & 0.332 & 0.347 \\
& Qwen3 Max & 0.427 & 0.283 & 0.317 & 0.552 & 0.422 & 0.441 & 0.492 & 0.342 & 0.372 & 0.466 & 0.323 & 0.353 & 0.498 & 0.323 & 0.360 & 0.487 & 0.339 & 0.369 \\
\midrule
\multirow{6}{*}{Hybrid-Review}
& Claude Haiku 4.5 & 0.252 & 0.233 & 0.220 & 0.251 & 0.289 & 0.246 & 0.288 & 0.283 & 0.260 & 0.249 & 0.252 & 0.227 & 0.313 & 0.266 & 0.260 & 0.270 & 0.265 & 0.243 \\
& DeepSeek V3.2 & 0.233 & 0.242 & 0.218 & 0.335 & 0.419 & 0.343 & 0.310 & 0.334 & 0.293 & 0.295 & 0.346 & 0.292 & 0.366 & 0.354 & 0.327 & 0.308 & 0.339 & 0.295 \\
& Gemini 3 Flash & 0.257 & 0.119 & 0.151 & 0.260 & 0.122 & 0.154 & 0.308 & 0.155 & 0.192 & 0.258 & 0.127 & 0.160 & 0.291 & 0.134 & 0.170 & 0.275 & 0.131 & 0.165 \\
& GPT-5.2 & 0.255 & 0.281 & 0.236 & 0.313 & 0.437 & 0.328 & 0.299 & 0.348 & 0.285 & 0.248 & 0.328 & 0.250 & 0.314 & 0.321 & 0.275 & 0.286 & 0.343 & 0.275 \\
& Kimi K2 & 0.205 & 0.113 & 0.135 & 0.298 & 0.185 & 0.214 & 0.272 & 0.151 & 0.180 & 0.233 & 0.141 & 0.165 & 0.253 & 0.137 & 0.166 & 0.252 & 0.145 & 0.172 \\
& Qwen3 Max & 0.196 & 0.161 & 0.161 & 0.273 & 0.277 & 0.253 & 0.249 & 0.216 & 0.210 & 0.220 & 0.203 & 0.190 & 0.335 & 0.250 & 0.260 & 0.254 & 0.221 & 0.215 \\
\bottomrule
\end{tabular}

\vspace{-10pt}
\end{table*}

\subsubsection{Performance of PR-Level ACR Baselines}

We further evaluate representative PR-level ACR baselines under the defect detection setting of \MCRbench, with results reported in Table~\ref{tab:acr_tool_results}. Overall, both baselines achieve limited performance, with generally low F1 scores and consistently low recall. This indicates that existing PR-level review baselines still struggle to capture defects whose evidence evolves across review discussions and code revisions.

\textbf{Comparison between ACR baselines.}
PR-Agent generally achieves higher overall F1 than Hybrid-Review, likely because their designs provide different types of evidence for defect detection. PR-Agent directly analyzes PR diffs and related PR context through an agent-based review workflow, while Hybrid-Review relies on static-analysis warnings to guide review generation. Such warnings can help identify certain local code issues, but may provide limited evidence for defects whose identification depends on evolving discussions, attempted fixes, or cross-round state changes, leaving Hybrid-Review more constrained by low recall in \MCRbench.
\textbf{Comparison with directly prompted LLMs.}
Both ACR baselines generally obtain lower performance than the directly prompted base LLMs in Table~\ref{tab:overall_performance}. This suggests that current ACR pipelines do not consistently turn their PR-level review workflows into better multi-round defect detection. This limitation is likely related to their original design goals: these pipelines are primarily designed for single-pass PR review comment generation, where the focus is on producing useful feedback from the current diff and local PR context, whereas \MCRbench requires models to connect review discussions, code revisions, and defect lifecycle states across rounds. As a result, pipeline-level context selection, prompting, or intermediate processing may fail to preserve all defect-relevant signals.


\subsubsection{Defect State Tracking Capability}
Table~\ref{tab:state_accuracy_overall} reports the accuracy of defect lifecycle state prediction for each LLM, which is calculated exclusively on the subset of correctly identified defects (True Positives). This is because predicting the lifecycle state is only meaningful when the defect itself has been successfully identified by the LLM. 

\begin{wraptable}{r}{0.5\textwidth}
\centering
\tiny
\vspace{-10pt}
\setlength{\tabcolsep}{3pt}
\renewcommand{\arraystretch}{1.15}
\caption{Accuracy of different models across programming languages and overall performance.}
\label{tab:state_accuracy_overall}
\vspace{-10pt}
\begin{tabular}{lcccccc}
\toprule
\textbf{Model} & \textbf{Py} & \textbf{Java} & \textbf{JS} & \textbf{TS} & \textbf{C\#} & \textbf{Overall} \\
\midrule
Claude Haiku 4.5 & \textbf{80.74\%} & \textbf{82.77\%} & \textbf{79.76\%} & \textbf{78.36\%} & \textbf{75.84\%} & \textbf{79.69\%} \\
DeepSeek V3.2   & 72.53\% & 77.24\% & 73.12\% & 68.59\% & 69.34\% & 72.60\% \\
GLM-4.7         & 61.39\% & 73.68\% & 61.69\% & 60.11\% & 61.25\% & 64.49\% \\
GPT-5.2         & 70.40\% & 75.31\% & 66.60\% & 70.55\% & 71.04\% & 71.23\% \\
Kimi K2         & 43.57\% & 48.97\% & 50.66\% & 45.00\% & 41.26\% & 45.95\% \\
Qwen3 Max       & 47.26\% & 45.13\% & 44.41\% & 44.93\% & 40.82\% & 44.34\% \\
\bottomrule
\end{tabular}

\vspace{-10pt}

\end{wraptable}

From an LLM-centric perspective, Claude Haiku~4.5 achieves the strongest performance in defect lifecycle state prediction across all programming languages, with an overall accuracy close to 80\% and relatively stable results across languages. GPT-5.2 and DeepSeek~V3.2 follow, both reaching overall accuracies around 70\%, but showing greater variation across different languages. In contrast, the remaining LLMs exhibit substantially lower accuracy on this task, reflecting a notable gap in multi-round state reasoning capability.

From a language-centric perspective, models generally achieve higher lifecycle state prediction accuracy on Python and Java than on JavaScript, TypeScript, and C\#. This pattern is consistent across LLMs, suggesting that state prediction performance varies with programming language settings. For languages such as TypeScript and C\#, most models struggle to accurately capture signals of cross-round state transitions, leading to noticeably lower prediction accuracy.

It is worth noting that even under the prerequisite of correct defect identification, LLM performance on lifecycle state prediction does not reach saturation. This observation indicates that lifecycle state identification is not a superficial classification task, but rather depends on a model’s integrated understanding of code changes, review comments, and historical context across multiple rounds, thereby placing higher demands on cross-round reasoning ability.


\begin{table}[htbp]
\vspace{-5pt}
\centering
\scriptsize
\setlength{\tabcolsep}{3pt}
\caption{Hit Rate of different LLMs across taxonomy categories.}
\label{tab:taxonomy_hit_rate_overall_row}
\vspace{-10pt}
\begin{tabular}{lccccccccccccc}
\toprule
\textbf{LLM} 
& \textbf{E.1.1} & \textbf{E.1.2} & \textbf{E.2} & \textbf{E.3.1} & \textbf{E.3.2} & \textbf{E.4}
& \textbf{F.1} & \textbf{F.2} & \textbf{F.3} & \textbf{F.4} & \textbf{F.5} & \textbf{F.6} & \textbf{F.7} \\
\midrule
Claude Haiku 4.5
& 0.5899 & 0.7033 & 0.5945 & 0.5685 & 0.5923 & 0.5439
& 0.5805 & 0.6510 & 0.6436 & 0.6526 & 0.5170 & 0.5370 & 0.5714 \\
DeepSeek V3.2
& 0.4165 & 0.4923 & 0.4377 & 0.4174 & 0.4097 & 0.3606
& 0.3702 & 0.4362 & 0.4078 & 0.4852 & 0.3693 & 0.3824 & 0.3333 \\
Gemini 3 Flash
& 0.5065 & 0.5793 & 0.4957 & 0.4515 & 0.5091 & 0.4798
& 0.5000 & 0.5346 & 0.5498 & 0.5899 & 0.4500 & 0.4300 & 0.4737 \\
GLM-4.7
& 0.3828 & 0.4361 & 0.4154 & 0.3619 & 0.3537 & 0.3589
& 0.3655 & 0.4428 & 0.3881 & 0.5123 & 0.4246 & 0.3512 & 0.4167 \\
GPT-5.2
& 0.5161 & 0.6000 & 0.5094 & 0.4785 & 0.5008 & 0.4735
& 0.5292 & 0.5820 & 0.6029 & 0.6483 & 0.5028 & 0.5243 & 0.4872 \\
Kimi K2
& 0.2807 & 0.3746 & 0.2061 & 0.1977 & 0.2748 & 0.2967
& 0.3552 & 0.4160 & 0.4363 & 0.4773 & 0.2924 & 0.3005 & 0.3714 \\
Qwen3 Max
& 0.3072 & 0.4034 & 0.2030 & 0.2582 & 0.2939 & 0.3516
& 0.4211 & 0.4311 & 0.4743 & 0.5369 & 0.2899 & 0.3074 & 0.2766 \\
\midrule
\textbf{Overall (Avg.)}
& \textbf{0.4285} & \textbf{0.5127} & \textbf{0.4088} & \textbf{0.3905} & \textbf{0.4192} & \textbf{0.4093}
& \textbf{0.4460} & \textbf{0.4991} & \textbf{0.5004} & \textbf{0.5575} & \textbf{0.4066} & \textbf{0.4047} & \textbf{0.4186} \\
\bottomrule
\end{tabular}
\vspace{-5pt}
\end{table}

\subsubsection{Deeper Analysis}

To gain more insights into LLMs' performance in multi-round code review scenarios, we conduct deeper analyses beyond overall performance metrics. Specifically, we examine performance from three perspectives: (1) LLMs’ defect identification capability across different defect categories, (2) LLM performance across different defect severity levels, and (3) typical error patterns in lifecycle state prediction.

\textbf{Hit Rate by Defect Category.}
Table~\ref{tab:taxonomy_hit_rate_overall_row} reports the hit rate of each LLM across different defect categories. Overall, LLMs' performance varies substantially across categories, and the average hit rate remains relatively low for many categories, indicating that defect identification difficulty is highly imbalanced across defect types in multi-round code review.

From Table~\ref{tab:taxonomy_hit_rate_overall_row}, several categories (e.g., E.1.2, F.3, and F.4) achieve relatively higher hit rates, with average values close to or above 0.50, which may reflect that defects in these categories are associated with clearer semantic cues in review discussions and are therefore comparatively easier for LLMs to recognize. In contrast, categories such as E.2, E.3.1, and several F subtypes exhibit noticeably lower average hit rates, pointing to ongoing challenges for LLMs in capturing implicit logic or cross-round evolution characteristics associated with these defects. Notably, these category-level performance differences are largely consistent across models: categories that are easier or harder to identify tend to follow similar rankings for most LLMs, suggesting that the observed disparities are more closely related to the semantic and structural properties of defect categories than to idiosyncratic behaviors of individual models.


\begin{wraptable}{r}{0.6\textwidth}
\centering
\scriptsize
\setlength{\tabcolsep}{3.8pt}
\caption{Hit rates of LLMs across defect severity levels.}
\label{tab:severity_hit_rate_overall}
\vspace{-10pt}
\begin{tabular}{lcccccc}
\toprule
\textbf{LLM}
& \textbf{Trivial} & \textbf{Minor} & \textbf{Normal}
& \textbf{Major} & \textbf{Blocker} & \textbf{Critical} \\
\midrule
Claude Haiku 4.5
& 0.5810 & 0.5609 & 0.5947 & 0.6707 & 0.6074 & 0.6607 \\
DeepSeek V3.2
& 0.4216 & 0.3970 & 0.4143 & 0.4424 & 0.3858 & 0.4808 \\
Gemini 3 Flash
& 0.4888 & 0.4815 & 0.4926 & 0.5761 & 0.5116 & 0.5472 \\
GLM-4.7
& 0.3910 & 0.3523 & 0.3892 & 0.4608 & 0.3876 & 0.5000 \\
GPT-5.2
& 0.4941 & 0.5069 & 0.5332 & 0.5964 & 0.4453 & 0.6667 \\
Kimi K2
& 0.2149 & 0.2455 & 0.3522 & 0.4227 & 0.3440 & 0.3800 \\
Qwen3 Max
& 0.2459 & 0.2857 & 0.3807 & 0.4439 & 0.2733 & 0.4247 \\
\midrule
\textbf{Overall (Avg.)}
& \textbf{0.4053} & \textbf{0.4043} & \textbf{0.4510}
& \textbf{0.5161} & \textbf{0.4221} & \textbf{0.5229} \\
\bottomrule
\end{tabular}
\vspace{-10pt}
\end{wraptable}

\textbf{Hit Rate by Defect Severity.}
As shown in Table~\ref{tab:severity_hit_rate_overall}, LLM performance varies across defect severity levels. In general, LLMs achieve higher hit rates on higher-severity defects (e.g., \emph{Major}, \emph{Blocker}, and \emph{Critical}) than on lower-severity ones (e.g., \emph{Minor} and \emph{Trivial}). From the overall averages (\emph{Overall Avg.}), the hit rates for \emph{Major} and \emph{Critical} defects exceed 0.5, whereas performance on \emph{Minor} and \emph{Trivial} defects remains substantially lower, a pattern that is largely consistent across different LLMs. Most models show relatively stable performance on higher-severity defects but exhibit a clear decline in hit rate as severity decreases, with the drop being most pronounced for \emph{Trivial} defects.

To further analyze this observation, we manually inspected several randomly sampled cases across different severity levels. We found that higher-severity defects usually have stronger behavioral impact and leave clearer signals in code changes, reviewer discussions, and subsequent revisions, making them easier for LLMs to identify. In contrast, lower-severity issues are often more subtle and depend on finer-grained semantic or convention-specific cues, making them easier to overlook.

\begin{wrapfigure}{r}{0.5\textwidth}
\centering
\includegraphics[width=\linewidth]{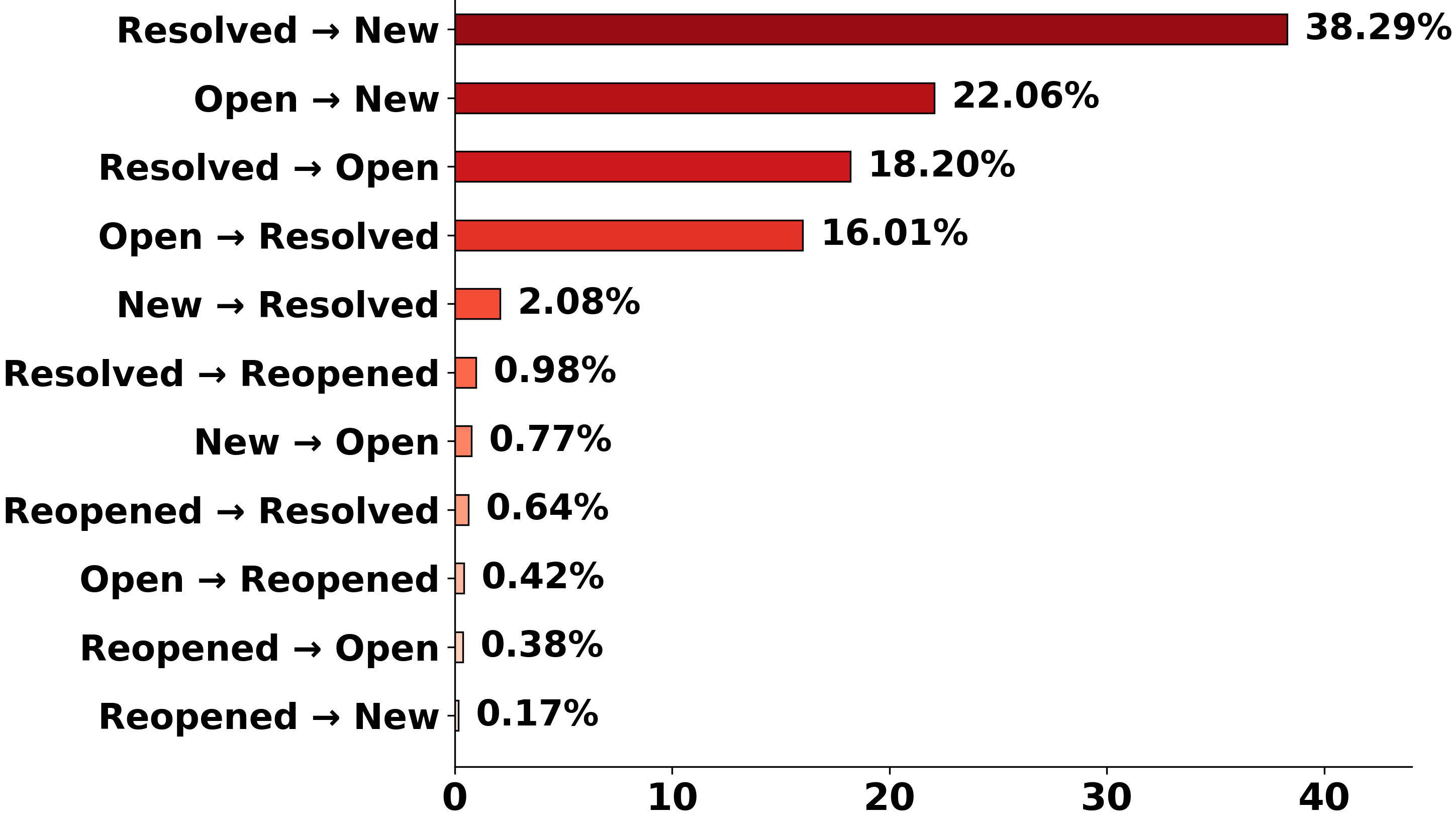}
\vspace{-15pt}
\caption{Distribution of LLMs' error patterns in multi-round defect lifecycle state tracking.}
\label{fig:error-patterns}
\end{wrapfigure}

\textbf{State Prediction Error Patterns. }
Figure~\ref{fig:error-patterns} illustrates the major error patterns in LLM predictions for the lifecycle state identification task. As shown in Figure~\ref{fig:error-patterns}, the most prominent state prediction error is confusing \emph{Resolved} defects as \emph{New} (38.29\%). This is followed by misclassifying \emph{Open} defects as \emph{New} (22.06\%) and predicting \emph{Resolved} defects as \emph{Open} (18.20\%). A comparable proportion of errors also arises from predicting \emph{Open} defects as \emph{Resolved} (16.01\%). 

As shown in Figure~\ref{fig:error-patterns}, the most prominent error pattern is erroneously inferring \emph{Resolved} defects as \emph{New}. This pattern suggests that models often fail to effectively leverage historical context across review rounds and tend to misinterpret already resolved defects as newly introduced defects, reflecting insufficient capability in remembering and integrating whether a defect has been fully addressed. Similarly, confusion from \emph{Open} to \emph{New} is relatively common, further indicating systematic difficulty in distinguishing persistent defects from newly introduced ones.

In addition, bidirectional confusion between \emph{Resolved} and \emph{Open} constitutes a substantial portion of errors. Such errors typically arise in cases where a defect has been partially addressed but review discussions are still ongoing, suggesting that models struggle to determine whether a defect has reached a fully resolved state. This highlights that lifecycle state prediction depends not only on the outcome of a single code modification, but also on a holistic understanding of review comments, subsequent feedback, and the adequacy of code changes.

In contrast, errors involving the \emph{Reopened} state occur at a lower frequency. This may be attributed to the lower prevalence of the \emph{Reopened} state in \MCRbench and its relatively clear semantic boundary, and also indicates that models are comparatively reliable at recognizing explicit reopening signals. However, when finer distinctions between \emph{Reopened} and other states are required, models may still produce incorrect predictions.


\begin{table}[t]
\centering
\footnotesize
\setlength{\tabcolsep}{7pt}
\caption{Average F1 score of different LLMs across increasing review rounds.}
\label{tab:round_f1_excl_r1}
\vspace{-10pt}
\begin{tabular}{lccccccccc}
\toprule
\textbf{LLM} & \textbf{R2} & \textbf{R3} & \textbf{R4} & \textbf{R5} & \textbf{R6} & \textbf{R7} & \textbf{R8} & \textbf{R9} & \textbf{R10} \\
\midrule
Claude Haiku 4.5 & \textbf{0.6495} & 0.6173 & 0.5359 & 0.5310 & 0.5313 & 0.4768 & 0.5227 & 0.5181 & 0.2857 \\
DeepSeek V3.2   & 0.5291 & 0.5559 & 0.5359 & 0.5148 & 0.4714 & 0.3964 & 0.3449 & 0.3667 & 0.3333 \\
Gemini 3 Flash  & 0.5983 & 0.5985 & 0.5554 & 0.5117 & 0.5371 & 0.5624 & 0.4910 & 0.3600 & 0.3333 \\
GLM-4.7         & 0.5025 & 0.5093 & 0.5220 & 0.5334 & 0.4839 & 0.4827 & 0.4967 & 0.5333 & 0.3333 \\
GPT-5.2         & 0.5995 & 0.5906 & 0.5659 & 0.5671 & 0.5067 & \textbf{0.6238} & 0.5000 & \textbf{0.6048} & \textbf{0.5000} \\
Kimi K2         & 0.3946 & 0.3463 & 0.3571 & 0.3450 & 0.3509 & 0.3356 & 0.4410 & 0.3467 & 0.3333 \\
Qwen3 Max       & 0.3927 & 0.3766 & 0.3580 & 0.3710 & 0.3624 & 0.3579 & 0.3654 & 0.3333 & 0.4444 \\
\bottomrule
\end{tabular}
\vspace{-10pt}
\end{table}

\vspace{-10pt}
\subsection{Performance Evolution across Rounds (RQ2)}

In this experiment, we investigate how interaction depth affects the performance of LLMs in multi-round code review scenarios, with a particular focus on whether models experience performance degradation under long-range interactions and whether they can maintain stable tracking of defects and their lifecycle states as contextual information continuously accumulates. Note that, Round~1 serves as the initial phase and inherently lacks the historical antecedents necessary to observe defect evolution. Consequently, to strictly align with our focus on multi-round dynamics and state transitions, we commence our performance analysis from Round~2 onward.

As shown in Table~\ref{tab:round_f1_excl_r1}, the average F1 scores of different LLMs across review rounds R2–R10 illustrate how performance evolves over successive interactions. As the number of review rounds increases, most LLMs exhibit varying degrees of performance degradation, indicating that long-range multi-round interactions substantially increase the difficulty of defect identification and state tracking. In earlier rounds (e.g., R2–R4), most LLMs maintain relatively stable performance; however, as contextual information accumulates, performance begins to fluctuate noticeably and, in some cases, degrades significantly in later rounds.

Additionally, performance under long-range interactions differs substantially across LLMs. Claude Haiku~4.5 achieves strong results in early review rounds but exhibits a gradual degradation as interaction depth increases, with a pronounced decline observed at R10. DeepSeek~V3.2, Gemini~3~Flash, and GLM-4.7 follow similar patterns, with F1 scores remaining at relatively low levels in later rounds. In contrast, GPT-5.2 demonstrates stronger stability in deeper interactions, maintaining comparatively higher F1 scores in later rounds such as R7 and R9. This pattern indicates that GPT-5.2 holds relative advantages in cross-round state tracking and long-context reasoning. Overall, the results show that mainstream LLMs face notable challenges in sustaining consistent performance in long-range, multi-round code review scenarios. As the number of review rounds increases, accumulated contextual noise and the dispersion of state-related cues substantially increase reasoning difficulty, thereby imposing greater demands on LLMs’ long-term memory and cross-round information integration capabilities.


\subsection{Root Cause Analysis (RQ3)}

\begin{table}[t]
\centering
\scriptsize
\renewcommand{\arraystretch}{1.25}
\setlength{\tabcolsep}{8pt}
\caption{Error taxonomy of False Positives (FP) and False Negatives (FN) in multi-round code review.}
\label{tab:error_taxonomy}
\vspace{-10pt}
\begin{tabular}{p{0.55cm}|p{2cm}|p{7.95cm}|p{0.7cm}}
\toprule
\textbf{Type} & \textbf{Error Category} & \textbf{Description} & \textbf{Prop.} \\
\midrule

\multirow[c]{5}{*}[-2.2em]{\centering \textbf{FP}}

& State--Temporal Misalignment 
& Failure to align defect states with code versions, causing already resolved defects to be repeatedly flagged as new. 
& 32.5\% \\
\cline{2-4}

& Over-reviewing 
& The model continues to produce review comments in the absence of real defects, misclassifying suggestions or speculative concerns as defects. 
& 27.8\% \\
\cline{2-4}

& Lack of Domain Knowledge 
& Insufficient understanding of project- or domain-specific conventions leads to reasonable but non-general implementations being misidentified as defects. 
& 22.4\% \\
\cline{2-4}

& Version Compatibility Blindness 
& Ignoring language or runtime version constraints and backward-compatibility requirements, the model suggests infeasible best-practice changes. 
& 12.1\% \\
\cline{2-4}

& Other 
& \centering --  
& 5.2\% \\
\midrule

\multirow[c]{6}{*}[-3.4em]{\centering \textbf{FN}}

& Long-range Dependency Miss 
& Failure to capture cross-file dependencies or downstream impacts of local changes results in missing real defects. 
& 23.4\% \\
\cline{2-4}

& Semantic Defect Blindness 
& Defects requiring semantic or specification-level reasoning are missed due to shallow pattern matching. 
& 22.3\% \\
\cline{2-4}

& Cross-round Defect Forgetting 
& The model fails to track unresolved defects across review rounds and stops mentioning them prematurely. 
& 25.1\% \\
\cline{2-4}

& Over-optimistic Fix Assumption 
& Partial fixes are incorrectly assumed to fully resolve defects, overlooking uncovered edge cases or paths. 
& 15.2\% \\
\cline{2-4}

& Sycophancy 
& Over-reliance on developer claims without sufficient code verification leads to missed defects. 
& 6.5\% \\
\cline{2-4}

& Other 
& \centering -- 
& 7.5\% \\
\bottomrule
\end{tabular}
\vspace{-10pt}
\end{table}

In multi-round code review scenarios, LLM errors can be broadly categorized into two types: \emph{False Positives} (FP) and \emph{False Negatives} (FN). These two error types represent distinct failure modes in multi-round code review—over-reviewing and defect omission—with substantially different causes and consequences. Specifically, FP errors occur when an LLM incorrectly predicts the presence of a defect or continues to flag a defect that does not exist or has already been resolved. While FN errors arise when an LLM fails to identify a defect that truly exists and remains unresolved, or prematurely stops tracking a defect in subsequent review rounds.

Given the distinct nature of FP and FN errors, we analyze them separately and adopt an open coding approach to construct a taxonomy of LLM failure causes in multi-round code review.
Specifically, we recruit four experienced programmers to manually analyze erroneous predictions produced by LLMs on \MCRbench using a two-stage open coding procedure, with FP and FN errors analyzed separately throughout. In the first stage, we randomly sample 10\% of FP and FN errors for analysis. The annotators independently examine the corresponding code changes, review discussions, and defect lifecycle information, assign descriptive labels to potential root causes, and then consolidate semantically similar labels through group discussions to form preliminary FP and FN taxonomies. In the second stage, all error instances are categorized based on the initial taxonomy, with new categories introduced when novel patterns emerge and previously annotated cases revisited to ensure consistency.

\textbf{FP Error Analysis.}
Table~\ref{tab:error_taxonomy} shows that FP errors mainly arise from several failure modes, among which \emph{State–Temporal Misalignment} constitutes the largest proportion (32.5\%) and represents the most critical root cause. This failure stems from the LLM’s inability to correctly align defect states with code review process evolution: even after a defect is fixed in later commits, the LLM may continue to rely on early-round comments or outdated code and label the defect as \emph{New} or reintroduce it. Such errors severely disrupt defect lifecycle tracking in multi-round settings, leading to repeated counting of the same defect and cascading misjudgments in later rounds, and reflect fundamental limitations in cross-round information integration and temporal reasoning.

Beyond state–temporal misalignment, the second prominent FP root cause is \emph{Over-reviewing} (27.8\%), which stems not from cross-round memory failures but from biased defect judgment. In the absence of sufficient evidence, LLMs tend to over-generate review comments or advisory suggestions as actual defects, reflecting difficulties in distinguishing optional improvements from mandatory fixes and in regulating review strictness. In addition, \emph{Lack of Domain Knowledge} (22.4\%) and \emph{Version Compatibility Blindness} (12.1\%) constitute another source of FP errors, as limited awareness of project-specific conventions or version constraints leads LLMs to flag context-dependent implementations or suggest infeasible changes. Overall, FP errors are dominated by over-reviewing under insufficient evidence and limited contextual understanding, highlighting systemic limitations in review decision boundaries and multi-round state maintenance.


\textbf{FN Error Analysis.}
As shown in Table~\ref{tab:error_taxonomy}, FN errors mainly stem from LLMs’ difficulties in maintaining consistent defect awareness across multi-round code review. Among the identified causes, \emph{Cross-round Defect Forgetting} is the most prevalent (25.1\%), where LLMs fail to continuously track unresolved defects across successive rounds. Defects identified early but not fully fixed are often prematurely dropped as attention shifts to newly introduced changes or discussions. This pattern highlights fundamental limitations in long-range state maintenance and helps explain the increasing miss rates observed as contextual information accumulates.

Beyond cross-round defect forgetting, FN errors are primarily driven by limitations in capturing deeper semantic and long-range dependencies. \emph{Long-range Dependency Miss} (23.4\%) reflects difficulties in reasoning about cross-file interactions and system-level effects, while \emph{Semantic Defect Blindness} (22.3\%) indicates failures to identify defects that require semantic or specification-level reasoning beyond surface patterns. In addition, \emph{Over-optimistic Fix Assumption} (15.2\%) reveals a bias toward prematurely judging defects as resolved based on partial changes or developer responses, which further amplifies missed detections in later rounds. Finally, \emph{Sycophancy} (6.5\%), though less frequent, highlights a tendency for LLMs to overly trust developer statements without sufficient code verification. Collectively, these error types underscore fundamental limitations in program semantic modeling, long-range reasoning, and independent verification, which hinder reliable defect identification in multi-round code review.


\subsection{Comment Quality (RQ4)}

Beyond defect detection and defect lifecycle state tracking, we further assess the quality of generated review comments. Specifically, we adopt ClearCRC~\cite{chen2025understanding}, a practitioner-oriented comment quality analysis framework derived from developers' expectations of clear code review comments. ClearCRC evaluates each generated review comment along three dimensions: \emph{Relevance}, which measures whether a comment is well aligned with the code change and review issue under discussion; \emph{Informativeness}, which measures whether the comment provides sufficiently concrete and useful information; and \emph{Expression}, which measures whether the comment is clearly and understandably phrased. Based on ClearCRC, we evaluate the quality of review comments generated by pure LLM prompting and ACR baselines.

\begin{table}[htbp]
\vspace{-5pt}

\centering
\footnotesize
\setlength{\tabcolsep}{3pt}
\renewcommand{\arraystretch}{1.0}
\caption{ClearCRC-based quality evaluation results for generated review comments.}
\label{tab:clearcrc_methods}
\vspace{-10pt}
\begin{tabular}{l rccl rccl rccl}
\toprule
\multicolumn{1}{c}{\multirow{2}{*}{\textbf{Model}}}
& \multicolumn{4}{c}{\textbf{Pure LLM}}
& \multicolumn{4}{c}{\textbf{PR-Agent}}
& \multicolumn{4}{c}{\textbf{Hybrid-Review}} \\
\cmidrule(lr){2-5}
\cmidrule(lr){6-9}
\cmidrule(lr){10-13}
& \textbf{Rel.} & \textbf{Info.} & \textbf{Expr.} & \textbf{Avg.}
& \textbf{Rel.} & \textbf{Info.} & \textbf{Expr.} & \textbf{Avg.}
& \textbf{Rel.} & \textbf{Info.} & \textbf{Expr.} & \textbf{Avg.} \\
\midrule
Claude Haiku 4.5 & 0.7215 & 0.4329 & 0.9909 & 0.7151 & 0.6887 & 0.7076 & 0.8194 & 0.7386 & 0.3559 & 0.3828 & 0.6445 & 0.4611 \\
DeepSeek V3.2    & 0.8891 & 0.5941 & 0.9553 & 0.8128 & 0.6624 & 0.6712 & 0.8684 & 0.7340 & 0.4654 & 0.4597 & 0.6599 & 0.5283 \\
Gemini 3 Flash   & 0.8552 & 0.6407 & 0.7977 & 0.7646 & 0.3117 & 0.2175 & 0.5345 & 0.3546 & 0.2649 & 0.1659 & 0.4717 & 0.3008 \\
GPT-5.2          & 0.8149 & 0.3598 & 0.9728 & 0.7158 & 0.7536 & 0.9302 & 0.9592 & 0.8810 & 0.6304 & 0.7968 & 0.8804 & 0.7692 \\
Kimi K2          & 0.9216 & 0.5389 & 0.9467 & 0.8024 & 0.7773 & 0.6526 & 0.9236 & 0.7845 & 0.2940 & 0.2459 & 0.5274 & 0.3557 \\
Qwen3 Max        & 0.8656 & 0.4795 & 0.9452 & 0.7634 & 0.7756 & 0.7379 & 0.8836 & 0.7990 & 0.3711 & 0.3558 & 0.5949 & 0.4406 \\
\bottomrule
\end{tabular}
\vspace{-10pt}

\end{table}

Table~\ref{tab:clearcrc_methods} reports the ClearCRC-based quality evaluation results for generated review comments. In the table, \emph{Rel.}, \emph{Info.}, and \emph{Expr.} denote \emph{Relevance}, \emph{Informativeness}, and \emph{Expression}, respectively, and \emph{Avg.} denotes their average. Overall, comment quality differs noticeably across generation settings and backbone models. Pure LLM prompting and PR-Agent generally achieve higher average scores than Hybrid-Review, while the best-performing setting varies by backbone model: pure LLM prompting performs strongly for DeepSeek V3.2 and Kimi K2, whereas PR-Agent achieves the highest averages for GPT-5.2 and Qwen3 Max.

At the dimension level, different settings show distinct quality profiles. Pure LLM prompting often obtains high \emph{Expression} scores but more uneven \emph{Informativeness} scores, suggesting fluent but not always sufficiently informative comments. PR-Agent tends to be more balanced for stronger backbones, especially in \emph{Informativeness}, while Hybrid-Review receives lower scores in most settings. These results indicate that comment quality and defect identification accuracy capture complementary aspects of automated code review, since higher-quality comments do not necessarily imply stronger defect coverage.


\section{Discussion}
\subsection{Potential Impacts}

\textbf{Benchmarking Multi-round Dynamics.}
\MCRbench provides a state-aware evaluation framework for modeling defect evolution across review rounds. By moving beyond single-round settings, it enables the assessment of cross-round defect identification and lifecycle tracking in more realistic multi-round review scenarios.

\textbf{Implications for Future Automated Code Review Research.}
Our findings suggest that future automated review methods should fully consider multi-round review scenarios during design. In particular, the performance drop across later rounds and frequent errors such as defect forgetting and temporal misalignment highlight the need for cross-round memory, historical context grounding, and lifecycle-aware state tracking~\cite{zhong2023memorybank}. Moreover, the ACR baseline and ClearCRC results indicate that pipeline complexity or fluent comments do not necessarily lead to stronger defect detection, suggesting that future research should jointly consider defect identification, state tracking, and comment quality.

\subsection{Threats to Validity}

\textbf{Internal Threats.}
Internal threats mainly relate to LLM selection, consistency filtering, and human annotation. First, the evaluated LLMs are limited in number. Although we include representative closed-source and open-weight LLMs, the results may not generalize to all existing or future LLMs. Second, consistency filtering may exclude some particularly difficult instances; however, we adopt this design to improve annotation reliability, and the modest overall performance of current models on \MCRbench suggests that the retained benchmark remains challenging in practice. Third, human annotation may involve subjective judgment. To mitigate this threat, we adopt a rigorous cross-validation protocol with independent double annotation, arbitration by a third annotator when disagreements arise, and a high inter-annotator agreement measured by Cohen's kappa~\cite{cohen1960coefficient} (0.87), thereby reducing individual bias and improving annotation reliability.

\textbf{External Threats.}
External threats primarily concern the representativeness of real-world code review scenarios and the coverage of evaluation metrics. As proprietary enterprise repositories are generally not publicly accessible, \MCRbench may not fully capture all aspects of enterprise-scale development; however, we focus on large, actively maintained repositories that closely reflect industrial coding practices and collaboration workflows. Another potential threat relates to the choice of evaluation metrics. While our evaluation does not exhaustively include all metrics proposed in the literature, we adopt a representative set of widely used and empirically validated metrics. Through a pre-study, we verify that the selected evaluation criteria exhibit strong consistency with human judgments, providing empirical support for their soundness and effectiveness.

\section{Conclusion}

In this paper, we introduce \MCRbench, a new benchmark for real-world multi-round code review. By modeling cross-round defect identification and defect lifecycle state transitions from practical review workflows, \MCRbench enables principled evaluation of LLMs’ abilities in defect discovery and tracking under continuous review settings, addressing the limitations of prior single-round-focused studies. Accordingly, \MCRbench provides a more realistic evaluation foundation for code review research and facilitates the development and assessment of more effective multi-round, state-aware LLM-based review systems for practical software development workflows.

\begin{acks}
This work is supported by the National Natural Science Foundation of China (Grant No. 92582202, No. 62302534), and GMCC-SYSU Joint Lab for Smart Applications.
\end{acks}

\section{Data Availability}
To facilitate the replication study, we have released our code and the experimental data at~\url{https://github.com/DeepSoftwareAnalytics/MCR-bench}.

\bibliographystyle{ACM-Reference-Format}
\bibliography{ref}

\end{document}